\documentclass[%
superscriptaddress,
preprint,
 amsmath,amssymb,
 aps,
floatfix,
]{revtex4-1}

\usepackage{graphicx}
\usepackage{caption}
\usepackage{subcaption}
\DeclareGraphicsExtensions{.png,.pdf}
\usepackage{dcolumn}
\usepackage[compact]{titlesec}
\usepackage{bm}
\usepackage{xcolor}
\usepackage{qcircuit}
\usepackage{braket}
\usepackage{float}
\usepackage{algorithm}
\usepackage{algorithmicx}
\usepackage{algpseudocode}
\usepackage{mathtools}

\usepackage{multirow}
\usepackage[normalem]{ulem}
\usepackage{hyperref}

\renewcommand{\figureautorefname}{Figure~\negthinspace}

\renewcommand{\tableautorefname}{Table~\negthinspace}
\renewcommand{\sectionautorefname}{Section~\negthinspace}

\begin{document}

\preprint{BNL}

\title{Decoding surface codes with deep reinforcement learning and probabilistic policy reuse}

\author{Elisha Siddiqui Matekole\footnote{Current affiliation: Riverlane}} %
 \email{esmatekole1@gmail.com}
\affiliation{%
 Computational Science Initiative, Brookhaven National Laboratory, Upton, NY 11973, USA
}%

\author{Esther Ye}
\email{estherye@bu.edu}
\affiliation{%
 Computational Science Initiative, Brookhaven National Laboratory, Upton, NY 11973, USA
}%
\affiliation{Department of Physics, Boston University, Boston, MA 02215, USA}

\author{Ramya Iyer}
\email{ramya.iyer@stanford.edu}
\affiliation{%
 Computational Science Initiative, Brookhaven National Laboratory, Upton, NY 11973, USA
}%
\affiliation{Stanford University, Stanford, CA 94305, USA}

\author{Samuel Yen-Chi Chen}
\affiliation{%
 Computational Science Initiative, Brookhaven National Laboratory, Upton, NY 11973, USA
}%




\date{\today}

\begin{abstract}

Quantum computing (QC) promises significant advantages on certain hard computational tasks over classical computers. However, current quantum hardware, also known as noisy intermediate-scale quantum computers (NISQ), are still 
unable to carry out computations faithfully mainly because of the lack of quantum error correction (QEC) capability. 
A significant amount of theoretical studies have provided various 
types of QEC codes; one of the notable topological codes is the surface code, and its features, such as 
the requirement of only nearest-neighboring two-qubit control gates and a large error threshold, 
make it a leading candidate for scalable quantum computation. 
Recent developments of machine learning (ML)-based techniques 
especially the reinforcement learning (RL) methods have been applied to the decoding problem and have already made certain progress. Nevertheless, the device noise pattern may change over time, making trained decoder models ineffective.
In this paper, we propose a continual reinforcement learning method to address these decoding challenges. Specifically, we implement double deep $Q$-learning with probabilistic policy reuse (DDQN-PPR) model to learn surface code decoding strategies for quantum environments with varying noise patterns. 
%
Through numerical simulations, we show that the proposed DDQN-PPR model can significantly reduce the computational complexity. 
Moreover, increasing the number of trained policies can further improve the agent's performance. 
Our results open a way to build more capable RL agents which can leverage previously gained knowledge to tackle QEC challenges.
\end{abstract}

\maketitle


\section{\label{sec:Introduction}Introduction}
Quantum computing (QC) promises superior capabilities in solving certain hard computational tasks which classical computers cannot solve in a reasonable time \cite{harrow2017quantum, nielsen2002quantum}. It has been shown theoretically that quantum computers can be used to factorize large numbers and, hence, 
break state-of-the-art public key cryptography systems \{almost\} exponentially \{superpolynomially\} 
faster than classical computers~\cite{shor1994algorithms}, and perform unstructured database searches with quadratic speedup~\cite{grover1996fast}. 
Such powerful algorithms require a quantum device with many qubits and a deep quantum circuit to carry out meaningful applications. 
Due to the stringent requirement on coherence, a vital necessity lies in keeping qubits coherent during operation when they are susceptible to noise and error-prone.
As opposed to classical error correction methods dealing mainly  with bit-flip errors~\cite{chang2006introduction}, quantum computers need to account for multiple types of errors, such as internal phase-flip errors, in addition to the bit-flip errors \cite{gottesman2002introduction, jayashankar2022quantum}.
Consequently, a small phase-angle error, or generally unitary error, close to the identity 
in a qubit could accumulate as multiple iterations take place. 
One way to address these and achieve the goal of a large functioning quantum device is to implement \emph{quantum error correction} (QEC)~\cite{shor1995scheme,steane1996error}.
The properties of quantum information pose challenges when developing QEC methods. 
The no-cloning theorem prevents us from directly duplicating an unknown quantum state~\cite{wooters1982nocloning}, and earlier works  of Shor~\cite{shor1995scheme} and Steane~\cite{steane1996error}, 
as well as others \cite{lidar2013quantum}, proposed ways to overcome this by encoding the logical states over multiple physical qubits in certain forms of entanglement. 

Most of the QEC decoding strategies are specific for a particular code realization. Moreover, it is challenging to implement the QEC strategies (including encoding and syndrome measurement) on the quantum hardware because it is very difficult to generate a precise model to capture all the dynamics of imperfect quantum hardware. Machine learning algorithms offer a solution to circumvent these issues and have a huge potential to outperform the current error correction strategies. Neural networks (NN) have been explored to enable the learning of error correction strategies for NISQ devices. It has been shown that with sufficient training, ML-based QEC can outperform conventional error-correction strategies \cite{Florian2018}, however, these techniques may not scale well as the number of qubits increases exponentially. 

NN-based decoders (also known as neural decoders) for topological codes can help to simplify the complex decoding algorithms to deduce the error syndrome as the size of quantum systems increases. Recently, stochastic NNs using the Boltzmann machine model have been proposed to realize neural decoders for topological codes. These decoders can accommodate different architectures and noise models. The NN learns decoding strategies by directly accessing the raw data obtained from measurements done on the quantum hardware \cite{Torlai2017}.
Another non-trivial task related to the error syndrome is to be able to predict the probability distribution of the errors for a QEC code. In \cite{krastanov2017deep}, NNs were deployed to encode this information.  
Even though this method is not scalable due to the probabilistic sampling, it paves the way to use more complex NN that can provide faster decoding as well as higher threshold values. Preliminary work on faster decoding was explored in \cite{varsamopoulos2017decoding,bhoumik2021efficient} for the case of a small surface code. 

Reinforcement learning (RL) is a machine learning method that learns how to make decisions via trial and error. Recent advances in deep neural networks (DNN) further empower RL capabilities and have already shown superhuman performance in solving complex sequential decision-making challenges such as playing the game of Go~\cite{silver2018general}. RL has been applied
to 
quantum control~\cite{sivak2022model, bukov2018reinforcement, niu2019universal}, quantum architecture search \cite{fosel2021quantum, kuo2021quantum}, and quantum error correction \cite{nautrup2019optimizing, andreasson2019quantum}. However, there is a concerning issue for RL, which is that classical simulation of training an RL agent is computationally expensive 
and time-consuming. Even worse, trained RL agents can only perform well when working on the task specified in their training and they cannot usually be generalized to different tasks even when the tasks are pretty similar~\cite{cobbe2019quantifying}. This poses a challenge when dealing with a quantum computing environment because the device noise changes over time \cite{proctor2020detecting}. To address this challenge, we make an addition to the traditional RL method to improve its performance in different quantum noise environments.

%
In this paper, we propose a continual RL approach to tackle the decoding problem with changing noise. 
%
Specifically, we construct a double deep $Q$-learning network (DDQN) agent equipped with probabilistic policy reuse (PPR) algorithm to improve the learning efficiency via utilizing previously learned knowledge. We show that the proposed framework can significantly reduce the required training episodes compared to training from scratch. In addition, we show that by increasing the number of trained policies in the policy library, the RL training can be better than training it with a smaller policy library.
The paper is organized as follows. In \sectionautorefname{\ref{sec:QuantumErrorCorrection}}, we provide a background on QEC and the basic idea of surface codes. In \sectionautorefname{\ref{sec:DeepReinforcementLearning}}, we introduce the RL methods used in this work. In \sectionautorefname{\ref{sec:ProbabilisticPolicyReuse}}, we describe the probabilistic policy reuse (PPR) framework used to extend RL agents. In \sectionautorefname{\ref{sec: Method}}, we provide the details of the experimental setup such as the quantum simulation environment and the RL training parameters. We demonstrate the results in \sectionautorefname{\ref{sec:ExpAndResults}}. Finally, we discuss the results in \sectionautorefname{\ref{sec:Discussion}} and conclude the paper in \sectionautorefname{\ref{sec:Conclusion}}.
%


\section{\label{sec:QuantumErrorCorrection}Quantum Error Correction and Surface Codes} 
There has been tremendous development in finding physical systems that can be used as qubits to encode quantum information. Some examples include neutral atoms \cite{Henriet2020quantumcomputing}, NMR-spin qubits \cite{vandersypen2001experimental}, NV-centers \cite{chen2019universal}, photonics \cite{slussarenko2019photonic,knill2001scheme,bartolucci2021fusion}, superconducting qubits \cite{national2019quantum,rosenblum2018cnot, lardinois2020ibm, meng2021cloud}, trapped ions \cite{allen2017reconfigurable,bohnet2016quantum}. 

However, qubits are very sensitive to the noise from their environment,  implying that the physical qubits are not able to carry out reliable logical computation and therefore limit the potential application of quantum computing. For fault-tolerant quantum computing, these physical qubits are required to build logical qubits, through \textit{Quantum Error Correction} (QEC).

QEC uses several physical qubits to make a single \emph{logical qubit}. In classical error correction, we can simply duplicate the information and use multiple bits to store the information. However in QEC, due to the fundamental limitations of the \emph{no-cloning theorem}, we cannot copy the unknown quantum information directly, hence we cannot \textcolor{purple}{directly} apply what we do in the classical world. Scientists have discovered several ways to entangle the qubits so that the quantum information is distributed into a set of physical qubits such that when some of the qubits fail, we can still recover the information. 

The basic idea of a QEC scheme is to define how to encode the quantum information with multiple qubits and how to perform \emph{parity checks} to know what errors actually occur \cite{devitt2013quantum, lidar2013quantum, roffe2019quantum, gottesman2002introduction, gottesman2010introduction}. After collecting enough information, the corresponding recovery routines can be carried out to correct the quantum state. 
There are several approaches to implementing QEC codes in a fault-tolerant setting. QEC codes that use four data qubits have been demonstrated \cite{corcoles2015demonstration, takita2017experimental,linke2017fault,andersen2020repeated}, but they cannot correct the error identified.
Classical repetition codes have also been used to correct errors \cite{cory1998experimental,chiaverini2004realization,reed2012realization,riste2015detecting,gunther2021improving,Wootton2018,google2021exponential}, however trying to make them scalable for larger quantum systems is still an open research question. 
A leading approach to scalable quantum computing in the NISQ era is to use topological codes \cite{fowler2012surface}. A prominent topological code is the surface code as it has an error threshold of $1\%$, and is compatible with planar architecture \cite{Raussendorf2007}.

\subsection{Stabilizer Formalism}
Most codes are stabilizer codes; one measures stabilizer operators to obtain error syndromes and uses this information to correct errors. The stabilizer formalism is a mathematical framework to describe QEC schemes and was first introduced by Daniel Gottesman \cite{gottesman1997stabilizer, gottesman2010introduction}. A quantum state $\ket{\psi}$ is defined to be stabilized by some operator $M$ if it is a $+1$ eigenstate of $M$: $M\ket{\psi} = \ket{\psi}$.
%
%
An $N$-qubit stabilizer state $\ket{\psi}_{N}$ is defined by the $N$ generators of an abelian 
subgroup $\mathcal{G}$ of the $N$-qubit Pauli group $\mathcal{P}_{N}$,
\begin{equation}
\label{eqn:stabilizer_definition}
\left.\mathcal{G}=\left\{M^{i}\left|M^{i}\right| \psi\right\rangle=|\psi\rangle,\left[M^{i}, M^{j}\right]=0, \forall(i, j)\right\} \subset \mathcal{P}_{N}
\end{equation}
%
Consider the stabilizer group $\mathcal{S}$, the \emph{codespace} is the set of quantum states $\ket{\psi}$ which are simultaneous eigenvectors of $M \in \mathcal{S}$ with eigenvalue $+1$. If a certain error happens, meaning that a Pauli operator $X$ or $Z$ acted on the quantum state, then some of the parity qubits may report eigenvalue $-1$, indicating that there is an error. The measurement results of parity qubits can be used to find out which qubits have errors and corresponding recovery procedures can be applied.
\subsection{Surface Code}
%
%
\begin{figure}[htbp]
    \centering
    \scalebox{0.21}{
    \centering
    \includegraphics[trim={0 14cm 0 0},clip]{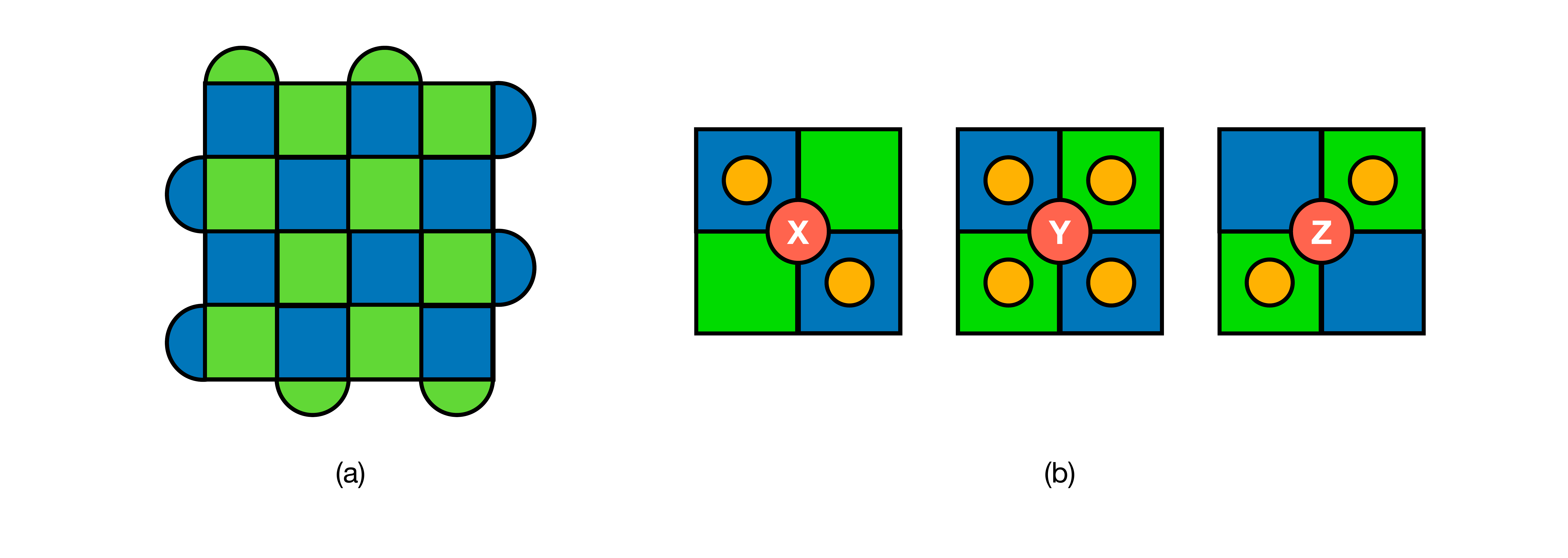}
    }
    \caption{{\bfseries Surface Code Diagram. } (a) A $d\times d$ square lattice ($d=5$) surface code diagram. The green (blue) plaquettes represent the independent parity checks for Pauli $X$ ($Z$) flips. Each vertex of the colored stabilizers contains a physical data qubit. (b) The $2 \times 2$ square lattice figures demonstrate the Pauli flips errors and the effect on the surrounding colored stabilizers. We can see that if there is a $X$ error, then two $Z$ parity checks will be in the eigenvalue $-1$. On the other hand, if there is a $Z$ error, then two $X$ parity checks will be in the eigenvalue $-1$. Since the $Y$ error is the combination of $X$ and $Z$ error, we can see that $Y$ error will make both $X$ and $Z$ parity checks to be $-1$.} 
    \label{Fig:surface}
\end{figure}
Various QEC frameworks have been developed under the stabilizer formalism. One of the most promising works is the \emph{surface code} \cite{bravyi1998quantum, fowler2012surface}. 
%

The surface code is a version of Kitaev's toric code \cite{kitaev2003fault, fujii2015quantum}, where the periodic boundary conditions are replaced by open boundary conditions \cite{bravyi1998quantum,freedman2001projective}. 
In this work, we consider the surface code used to encode a single qubit. In \figureautorefname{\ref{Fig:surface}} we show the $d\times d$ lattice, where each vertex represents a physical qubit. The independent parity checks for $X$ ($Z$) are represented by the green (blue) plaquettes and are modified at the boundary to act on three qubits. 
In order to measure the error without destroying the logical qubits, we need to perform projective measurements on ancilla qubits. These types of measurements are known as stabilizer measurements. The outcome of these ancilla qubit measurements constitutes the \emph{syndrome}, which informs us whether an error has occurred on the logical qubit.  


\subsection{\label{subsec:decoding}Decoding Scheme}
Syndrome measurements allow for the usage of a decoding scheme to i) process any information about errors that have occurred and ii) correct those errors appropriately. Depending on which error correction code is used, it is possible to obtain information about the existence, location, and type of error by using syndrome measurements. With this information, a decoder can then correct the errors that were detected and have the means to verify its success (through conducting subsequent syndrome measurements).

Since errors occur continuously over time, the decoding scheme can be treated as a process that also occurs continuously. The aim of such a scheme would still be to correct errors, with the added stipulation to do so for as long as possible (thus extending the \textit{lifetime} of the qubit). This has led to the development of decoders that successfully utilize machine learning as a means to continuously combat errors \cite{kim2020quantum,convy2022machine}. 
In our work, we utilize reinforcement learning to train our agent to act as a decoder for the surface code.
\section{\label{sec:DeepReinforcementLearning}Deep Reinforcement Learning}
\emph{Reinforcement learning} (RL) is a learning strategy where the training is based on learning from experience, and building strategies to solve the given problem. For a more detailed discussion on RL, we refer the readers to \cite{sutton2018reinforcement}. 
Generally, RL is comprised of two main elements: an \emph{agent} and an \emph{environment}. The \emph{agent} interacts with the \emph{environment} $\mathcal{E}$ over a collection of discrete time steps, where $\mathcal{E}$ contains the necessary information to describe the problem at hand. It contains the rules of the game, the set of all possible states $s \in S $ (observations), and feedback on the quality of the actions taken by the agent. At any time step $t$, the agent receives a state $s_{t}$ from the $\mathcal{E}$, and chooses an \emph{action}, $a_{t} \in \mathcal{A}$. This is done with respect to the agent’s \emph{policy} $\pi$, which maps $s_t$ to action $a_t$. The \emph{policy} $\pi$ is generally, stochastic in nature, such that for a given $s_{t}$, the action output is a probability distribution $\pi(a_{t}|s_{t})$. Now, an \emph{episode} in the training is defined as the agent starting with some random initial state $s_{0}$ and interacting with the $\mathcal{E}$ following the aforementioned process. At any time $t$, after executing the action $a_{t}$, the agent receives the state of the next time step $s_{t+1}$ and a scalar \emph{reward} $r_{t}$. This process continues until the agent satisfies a pre-defined stopping criterion or a terminal state. 

The reward depends on the state and the action taken. The agent gets a higher reward when it progresses towards the goal and a penalty when performing bad actions. Concretely speaking, the objective of any RL problem is to find an optimal policy $\pi^{*}$ that maximizes the rewards. The total discounted return from time step $t$ is defined as $R_t = \sum_{t'=t}^{T} \gamma^{t'-t} r_{t'}$, where $\gamma$ is the discount factor that lies in $(0,1]$. Here $\gamma$ is the parameter that controls how future rewards are weighted to the decision making function. When a large $\gamma$ is considered, the agent weighs the future reward more heavily. On the other hand, with a small $\gamma$, the agent weighs the immediate reward more.
The expected return for selecting an action $a$ in state $s$ based on policy $\pi$ is defined as the \emph{action-value function} or \emph{$Q$-value function} $Q^\pi (s,a) = \mathbb{E}[R_t|s_t = s, a]$. The optimal action value function $Q^*(s,a) = \max_{\pi} Q^\pi(s,a)$ gives a maximal action-value across all possible policies. The value of state $s$ under policy $\pi$, $V^\pi(s) = \mathbb{E}\left[R_t|s_t = s\right]$, is the agent's expected return by following policy $\pi$ from the state $s$. The RL algorithms which maximize the value function are called \emph{value-based} RL.

\subsection{\textit{Q}-Learning}
$Q$-learning \cite{sutton2018reinforcement} is one of the  most widely used model-free approaches in RL. In $Q$-learning, the agent learns the optimal action-value function and is an \emph{off-policy} algorithm. 
The learning begins by arbitrarily initializing the value function $Q^{\pi}(s,a) \forall s\in S, a\in \mathcal{A}$, typically stored in a table known as the $Q$-table. The estimates for $Q^{\pi}(s,a)$ are then progressively updated according to policy using the Bellman equation: 
\begin{align}
  Q\left(s_{t}, a_{t}\right) \leftarrow  Q\left(s_{t}, a_{t}\right)
  +\alpha\left[r_{t}+\gamma \max _{a} Q\left(s_{t+1}, a\right)-Q\left(s_{t}, a_{t}\right)\right].
\end{align}

\subsection{\label{subsec:DDQ}Double Deep \textit{Q}-Learning}
Although the previously explained method of $Q$-learning gives the optimal action-value function, it is not feasible for problems that require larger memory. For example, it would be very difficult to deal with problems with high dimensions of state $s$ or action $a$. In order to get around this memory requirement, neural networks (NN) are used to efficiently represent $Q^{\pi}(s,a) \forall s \in S, a \in \mathcal{A}$. This method of using NNs to learn $Q$-values is known as \emph{deep $Q$-learning} and the network is called a deep $Q$-network (DQN) \cite{mnih2015human}. 

In order to stabilize the DQN, we use \emph{experience replay} and an additional network known as the \emph{target network} \cite{mnih2015human}. In experience replay, the agent stores the experiences encountered during the episodes in a memory which stores the transition tuple, $\{s_{t}, a_{t}, r_{t}, s_{t+1}\}$. After gathering enough experiences, the agent randomly samples a batch of experiences, computes the loss, and updates the DQN parameters. Additionally, in order to reduce the correlation between the target and prediction, a clone of DQN, known as a \emph{target network} is used. The DQN parameters $\theta$ are updated at every iteration while the target network parameters $\theta^{-}$ are updated every few iterations. The DQN learning is done via minimizing the mean square error (MSE) loss function:
\begin{equation}
    L(\theta)=\mathbb{E}\left[\left(r_{t}+\gamma \max _{a^{\prime}} Q\left(s_{t+1}, a^{\prime} ; \theta^{-}\right)-Q\left(s_{t}, a_{t} ; \theta\right)\right)^{2}\right]
\end{equation}


In this work, we extend this to implement \emph{Double Deep $Q$-learning}, as sometimes DQN can overestimate the action-value function \cite{hasselt2016qlearning}. The idea behind of double deep $Q$-learning is to decompose the max operation in the target $y^{DQN}_{t} = r_{t}+\gamma \max _{a^{\prime}} Q\left(s_{t+1}, a^{\prime} ; \theta^{-}\right)$ into two separate operations: \emph{action selection} and \emph{action evaluation}. The action selection is based on the policy network, $\operatorname{argmax}_{a} Q\left(s_{t+1}, a ; \theta\right)$ and then the target network is used to evaluate the action, $Q\left(s_{t+1}, \operatorname{argmax}_{a} Q\left(s_{t+1}, a ; \theta\right), \theta^{-}\right)$. The DDQN target is now $y^{DDQN}_{t} = r_{t}+\gamma Q\left(s_{t+1}, \operatorname{argmax}_{a} Q\left(s_{t+1}, a ; \theta\right), \theta^{-}\right)$. The loss function $L(\theta)$ is therefore:

\begin{equation}
    L(\theta)=\mathbb{E}\left[\left(r_{t}+\gamma Q\left(s_{t+1}, \operatorname{argmax}_{a} Q\left(s_{t+1}, a ; \theta\right), \theta^{-}\right) -Q\left(s_{t}, a_{t} ; \theta\right)\right)^{2}\right]
\end{equation}

Then, $\theta$ is updated using the gradient descent method and every few iterations we update the target network $\theta^{-}\leftarrow\theta$.
%
\section{\label{sec:ProbabilisticPolicyReuse}Probabilistic Policy Reuse}
%
\begin{figure}[htbp]
    \centering
    \scalebox{0.25}{
    \centering
    \includegraphics{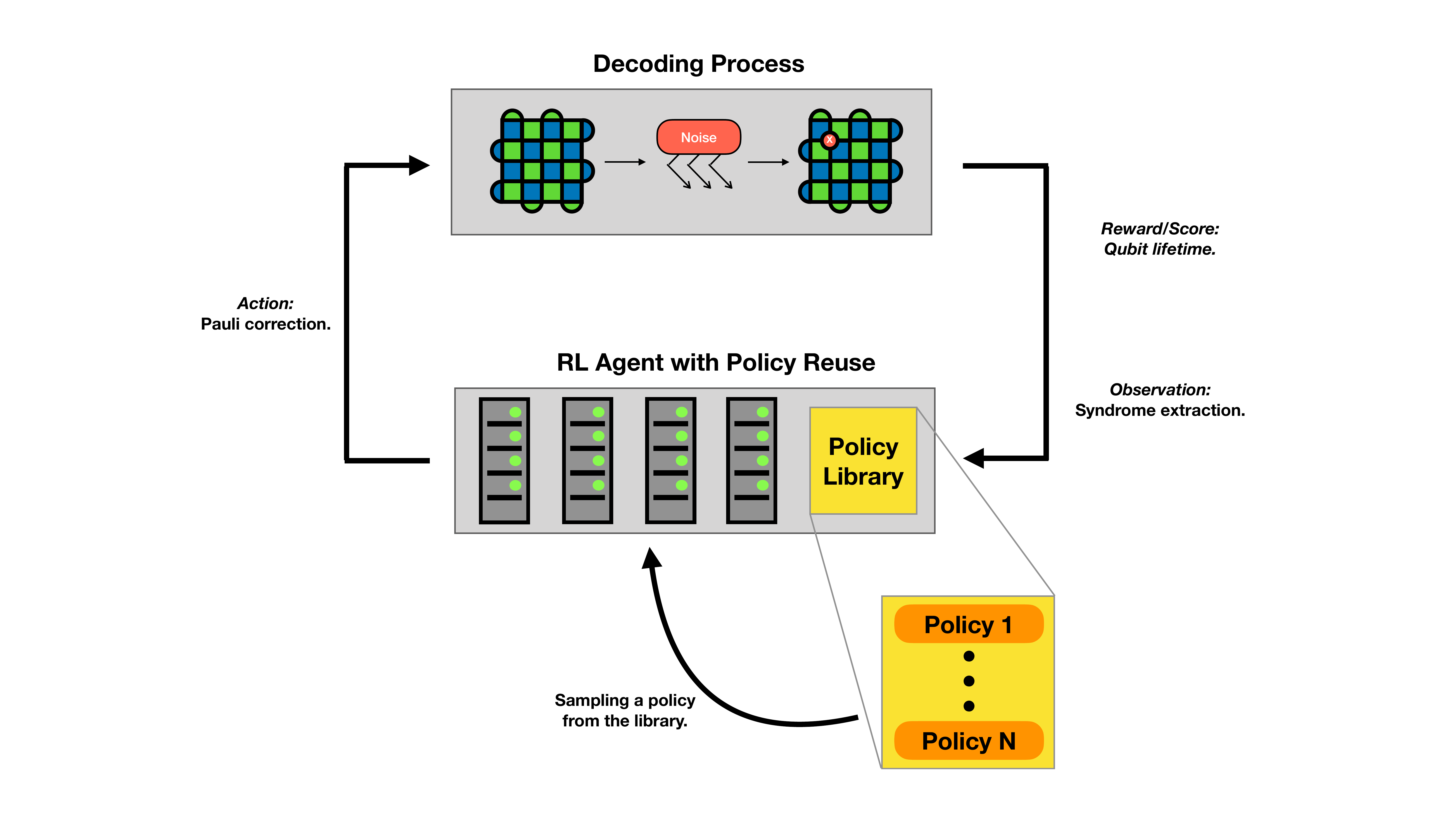}
    }
    \caption{{\bfseries Overview of Policy Reuse for Decoding.} Shown above is the cyclic process of the RL agent correcting errors on the surface code environment for as long as possible. Within the \textit{Decoding Process} box, we see that the surface code is constantly being subjected to noise, creating errors on the physical qubits stationed at each point on the surface code lattice. The errors are detected in the form of a syndrome extraction and fed to the RL agent. In the RL Agent box, we see that the agent has access to a policy library that has stored information previously gathered from other noise environments. The RL agent can leverage these policies when considering the best course of action to enact on the current surface code environment under consideration. }
    \label{Fig:ppr}
\end{figure}
As mentioned in \sectionautorefname{\ref{subsec:decoding}}, errors can occur continuously while the qubit is operating and ML-based decoders can be used to process and correct this stream of errors.
\textit{Probabilistic policy reuse (PPR)} is one implementation of \textit{continual learning}, which is a subset of machine learning specifying the ability to learn from information that is accumulated over time. To do this, a policy reuse algorithm in particular stores and makes use of its previously solved solutions (in the form of policies/models) to aid in its computation of an upcoming task. This method has been previously demonstrated with other architectures such as: utilizing PPR in conjunction with tabular $Q$-learning \cite{fernandez2006probabilistic} and deep $Q$-networks \cite{ye2021quantum}. \emph{Our algorithm will utilize this method with double deep $Q$-learning to conduct quantum error correction on the surface code.}

To further elaborate on how PPR operates, the previous policies are stored in a policy library, $L$, and sampled probabilistically at the beginning of each learning episode. This process acts outside, but in conjunction with, the DDQN framework discussed previously in \sectionautorefname{\ref{subsec:DDQ}} and utilizing the agent without any policies loaded into $L$ can be thought of as training the agent from scratch with just the DDQN. After each episode, a score is calculated to represent how well the policy performed. This score then allows for an update to the probabilistic weights associated with each policy stored in $L$, allowing for the better-performing policies to gain a greater probability of being chosen in the next episode. 

As stated earlier, the policy reuse process acts outside the DDQN as shown in \figureautorefname{\ref{Fig:ppr}}.
The policy to be examined for a given episode is sampled from the policy library according to the softmax equation:
\begin{equation}
    P(\Pi_{j}) = \frac{e^{\tau W_{j}}}{\sum_{p = 0}^{n}e^{\tau W_{p}}}
    \label{eqn:softmax}
\end{equation}
which outputs a probability vector $P$, with each vector component corresponding to the probability assigned to a given policy. $P$ is dependent on the current reward of each of the policies $W$, and a temperature parameter $\tau$.

The aim of incorporating PPR on top of the DDQN is to enhance the performance of the agent by leveraging information from previous solutions rather than starting from scratch when encountering a new task. This is especially relevant when considering how to correct errors such as the bit-flips described in  \sectionautorefname{ \ref{sec:QuantumErrorCorrection}} because the algorithm can recall policies that have previously solved similar error configurations to aid in its management of new errors. 
\section{\label{sec: Method}Method}
\subsection{Environment and Encoding of the agent}
The summary of the environment and the interaction of the agent in this section is based on the work presented in \cite{Sweke2020}. For more details, we encourage the readers to refer to this paper.
First, we outline the encoding of the environment state. The environment state $s_{t}$ at time $t$ is constructed by layering the action history slices and faulty syndrome slices together. 
In the \emph{environment} module, the $d\times d$ surface code lattice is encoded into a $(2d+1) \times (2d+1)$ binary lattice, enabling us to distinguish the $X$ flips and $Z$ flips. This kind of encoding allows us to input the state $s_{t}$ to a deep convolutional neural network (CNN). The function of the CNN is to isolate a particular patch of the lattice which contains information about the syndrome volume and previous actions. 
The output from the CNN layer is then fed to a feed-forward neural network. The final layer of this neural network encodes the $Q$-value, $Q(s_{t},a)$, where \emph{a} is the action. 
Next, we describe the parameterization of the $Q$-function using the deep CNN, which forms the deep $Q$-network. The deep $Q$-network is made up of convolutional layers, feed-forward neural layers, and a final layer that provides $Q$-values for the corrections of the different input states.

Finally, we describe the training of the agent as it interacts with the environment to yield the optimal $Q$-value. 
In this setting, the agent is trying to solve the decoding problem and converge on a strategy that maximizes the discounted cumulative reward. Every new episode is initiated by extracting a new syndrome volume (which is faulty due to the presence of measurement error with probability $p_{\rm{meas}}$) and resetting the action history to zeros. The extracted syndrome and the reset action history are then fed to the agent as described before 
The agent will then choose an action according to the exploration/exploitation strategy. Next, the chosen action is applied to the surface code and the error configuration is updated, as well as the action history. 

In order to determine the agent's reward, the perfect syndrome, with respect to the updated error configuration, is fed to the referee decoder. The referee decoder, given a perfect syndrome, is used to suggest corrections to move the current state back to the codespace. 
The referee decoder is a fast feed-forward NN trained using supervised learning based on \cite{varsamopoulos2017decoding,varsamopoulos2019neural,chamberland2018deep}. 
If the referee decoder is able to successfully decode the current syndrome, the agent remains alive; otherwise, the agent dies and the episode ends. If the action chosen by the agent restores the state back to the original state, the agent gets a reward of $1$, otherwise, it gets a reward of $0$. 
At the end of a complete learning episode, the cumulative number of correct actions the agent successfully to stabilize the qubit is extracted as the qubit lifetime.
The above process continues until the agent chooses identity as an action, which implies that the agent is confident that it has applied all the necessary actions to return the code to the desired initial state. This leads to a new syndrome volume, and a new state is constructed from the reset action history and the updated syndrome. The new state is fed to the agent and once again the episode continues as described previously, until the agent dies. 
In the following section, we use the above environment and training steps to implement our approach to the decoding problem via the probabilistic policy reuse. 
\setlength{\tabcolsep}{0.3em}
\begin{table}[H]
\centering
\begin{tabular}{|c|c|ccccc|}
\hline
 &       & \multicolumn{5}{c|}{Environments}                                                                                          \\ \hline
 &
   &
  \multicolumn{1}{c|}{\emph{Environment-0}} &
  \multicolumn{1}{c|}{\emph{Environment-1}} &
  \multicolumn{1}{c|}{\emph{Environment-2}} &
  \multicolumn{1}{c|}{\emph{Environment-3}} &
  \emph{Environment-4} \\ \hline
\multirow{5}{*}{\rotatebox[origin=c]{90}{error-probabilities}} &
  0.003 &
  \multicolumn{1}{c|}{Output} &
  \multicolumn{1}{c|}{Input} &
  \multicolumn{1}{c|}{Input} &
  \multicolumn{1}{c|}{Input} &
  Input \\ \cline{2-7} 
 & 0.005 & \multicolumn{1}{c|}{--} & \multicolumn{1}{c|}{Output} & \multicolumn{1}{c|}{Input}  & \multicolumn{1}{c|}{Input}  & Input  \\ \cline{2-7} 
 & 0.007 & \multicolumn{1}{c|}{--} & \multicolumn{1}{c|}{--}     & \multicolumn{1}{c|}{Output} & \multicolumn{1}{c|}{Input}  & Input  \\ \cline{2-7} 
 & 0.011 & \multicolumn{1}{c|}{--} & \multicolumn{1}{c|}{--}     & \multicolumn{1}{c|}{--}     & \multicolumn{1}{c|}{Output} & Input  \\ \cline{2-7} 
 & 0.015 & \multicolumn{1}{c|}{--} & \multicolumn{1}{c|}{--}     & \multicolumn{1}{c|}{--}     & \multicolumn{1}{c|}{--}     & Output \\ \hline
\end{tabular}
\caption{This table shows the sequence of different environments corresponding to the different error-probability $p_{\rm{err}}$. Based on the PPR method, previous policies are sampled from the policy library and used as inputs in the subsequent environments. The \emph{Environment-0}, which is the basis for all of the following experiments, will be trained from scratch with $p_{\rm{err}} = 0.003$ using DDQN.}. 
\label{table:1}
\end{table}
\subsection{\label{sec:ML model}Machine Learning Model}

\subsubsection{\label{sec: ExpSetup}\textbf{Experimental Setup}}
In this section, we implement deep reinforcement learning (DRL) and PPR to solve the decoding problem. We first train the agent from scratch for the given noise model and compare it with the cases where we use the PPR algorithm. Our goal is to demonstrate that the agent can solve the decoding problem, using previously solved models in different environments, faster than if it was to solve it from scratch. 
\subsubsection{\label{sec: Noise}\textbf{Noise model}}
We consider a simpler noise model where the $X$ and $Z$ errors are uncorrelated. Since correcting independent $X$ or $Z$ error is equivalent, it is sufficient to develop our algorithm to correct one type of error.
We test our agent's decoding ability for the bit-flip noise model ($X$-noise model) and observe the performance of our training algorithm under different noise environments.
\subsubsection{\label{sec: DDQN}\textbf{Deep Q-Network Setup}}
We use the Double Deep $Q$-Network (DDQN) (described in \sectionautorefname{\ref{subsec:DDQ}}) in all our simulations for both the training from scratch method and the PPR algorithm. For a given learning episode, the agent observes its current state at each time step $t$ and samples from an action set $\mathcal{A}$ of Pauli flips ($X$ or $Z$) and a special \textit{request new syndrome}, to choose one as the action which it believes will drive it towards the correct solution. In the current environment setting the agent is restricted to single-qubit operations.

We implement DDQN using the PyTorch packages. We kept the architecture of the neural network similar to the original paper \cite{Sweke2020}, using three layers of 2-dimensional CNN and two feed-forward linear networks. The kernel size and stride for the first CNN layer is $[3,2]$ respectively, and for the last two layers it is $[2,1]$ respectively.
The input to the DDQN is the state $s_{t}={s_{\rm{sv},t},~h_{t}}$, where $s_{\rm{sv},t}$ is the faulty syndrome volume and $h_{t}$ is the action history list.  
Next, we describe the implementation of our PPR algorithm. We first train the agent from scratch for a very small error-probability ($p_{\rm{meas}}=p_{\rm{phys}}=0.003$), \emph{Environment-0}. The probability of an error occurring on a single physical qubit is given by $p_{\rm{phys}}$ and the probability of error during syndrome measurement is given by $p_{\rm{meas}}$. In our simulations, we set $p_{\rm{phys}}=p_{\rm{meas}}=p_{\rm{err}}$. Therefore, in the rest of the work, we will use $p_{\rm{err}}$ to describe the error-probabilities. 
We then load this policy to solve for \emph{Environment-1}, where we increase the error-probability, making it a harder environment to solve. This makes it more complex for the agent to suppress the errors. Next, we combine the previous two models and load them to solve for the next environment which has a higher error-probability and is more difficult to solve. We show a schematic of our framework in \figureautorefname{\ref{Fig:ppr}}, and a table showing the sequence of the different environments used in building the policy library in \tableautorefname{\ref{table:1}}.

\begin{figure}[b]
    \centering

   
    \begin{subfigure}[b]{1\columnwidth}
     \includegraphics[angle=0,width=1\columnwidth]{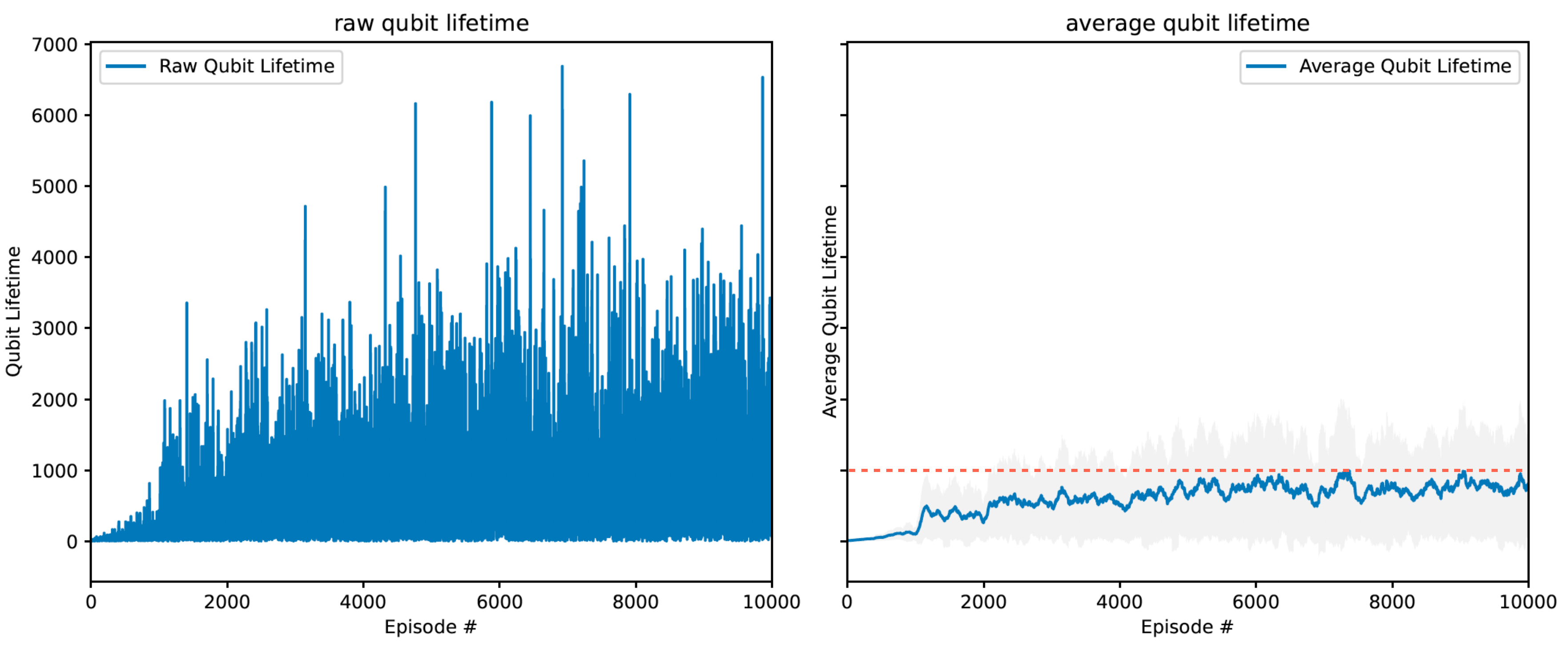}
    \end{subfigure}
   \caption{\emph{Environment-0} training from scratch simulation using DDQN, for a very small error-probability of $p_{\textrm{err}} = 0.003$}.
    \label{fig:Env0}
\end{figure}
\begin{figure}[b]
\centering
\begin{subfigure}[b]{1.\textwidth}
   \includegraphics[width=1\textwidth]{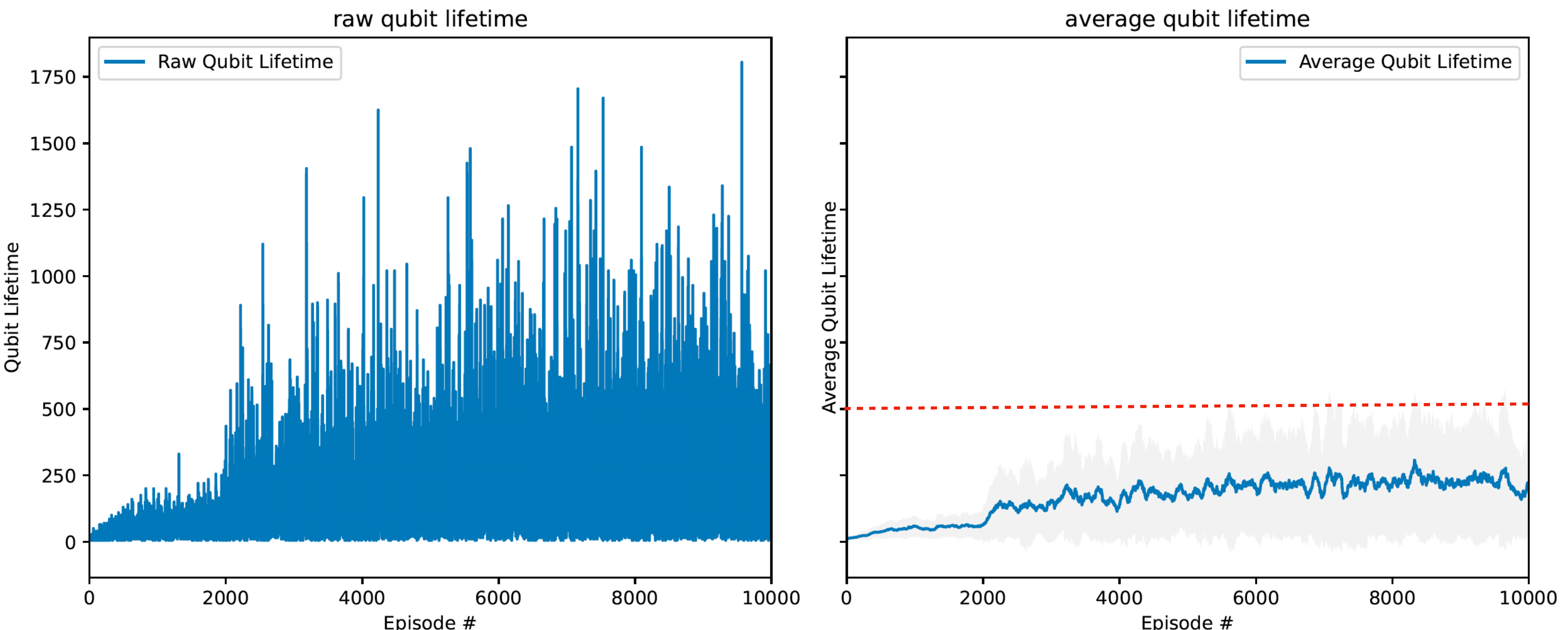}
   \caption{Training from scratch simulation for \emph{Environment-1} using DDQN.}
   \label{fig0.005scratch}
\end{subfigure}
\hfill
\begin{subfigure}[b]{1.\textwidth}
   \includegraphics[width=1\textwidth]{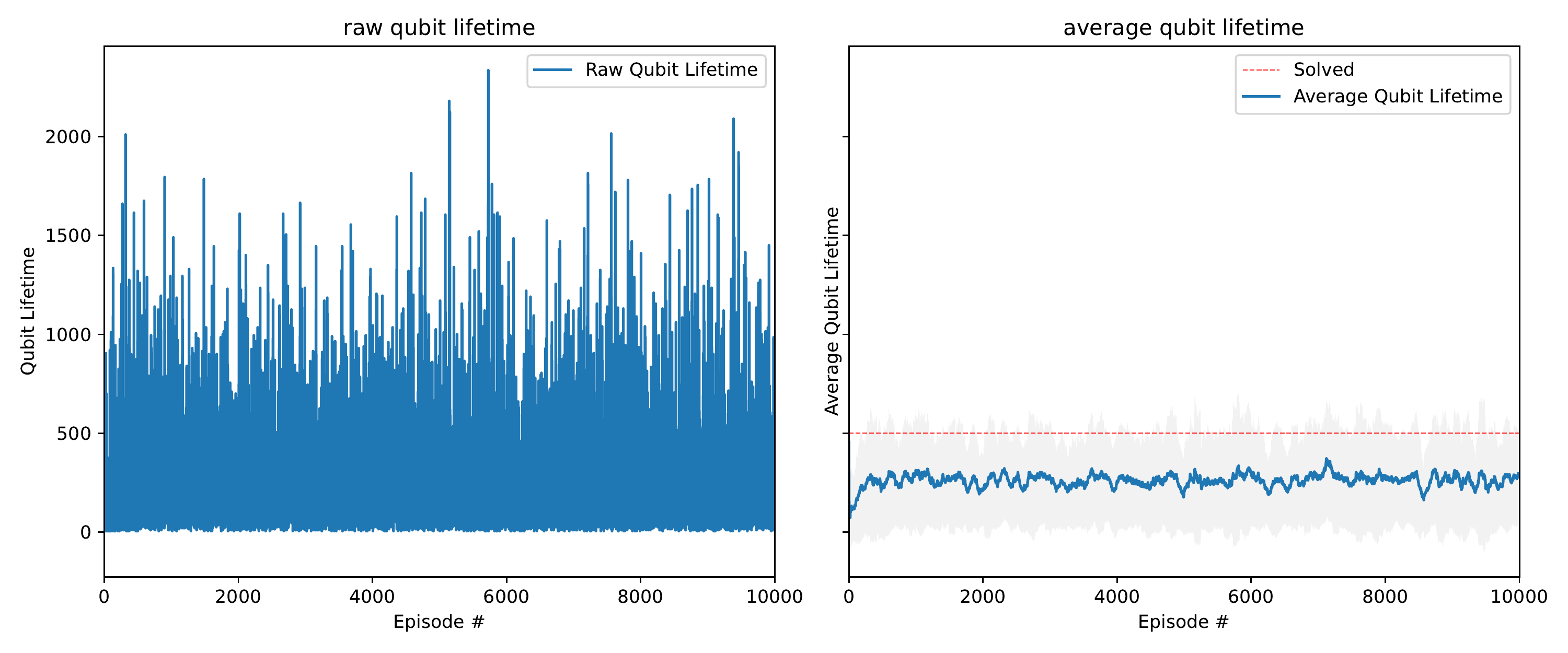}
   \caption{Policy reuse simulation for \emph{Environment-1}, using \emph{Environment-0} policy from the policy library.}
   \label{fig0.005_env1}
\end{subfigure}
 \caption{(Color online) Training from scratch and policy reuse results for \emph{Environment-1}: Single bit-flip error rate and a measurement error rate of 0.005. (a) Qubit lifetime as a function of episodes, when trained from scratch. We implement a DDQN neural network to train the agent.  (b) Qubit lifetime as a function of episodes, when using previously trained policy from \emph{Environment-0}, $p_{\textrm{err}}=0.003$. We use the qubit lifetime of 500 as the baseline to compare the results between the training from scratch and policy reuse methods respectively.  }
\label{fig0.005}
\end{figure}

\paragraph*{\textbf{\label{sec:PPRH}Probabilistic Policy Reuse Algorithm Hyper-parameters}} For the PPR algorithm, we assigned the following values to our hyper-parameters: For the temperature parameter $\tau$ (which is updated during each episode), we set the initial value to
be $0$. It increases incrementally with the number of episodes that have passed by $\delta\tau$, which we set as $0.01$. We set the number of total episodes $K$ to be $10000$ and the maximum number of steps in a given episode, $H$, was set to be $1000$. For the replay memory $D$ we initialized a capacity of $10000$ transition pairs.
The learning rate for the ADAM optimizer \cite{kingma2014adam} was set to $0.001$. Additionally, we have the following hyper-parameters for the $\pi$-exploration
algorithm: We assign the initial value (of the probability of following a previous policy)
to be $1$ and the value for $\nu$ (the decay factor of $\psi$ ) to be $0.95$. Also, we calculate the loss using the SMOOTH L1 LOSS function in PyTorch \cite{ren2015faster}.

\section{\label{sec:ExpAndResults}Experiments and Results}
\subsection{\emph{Environment-0}: error-probability=0.003 }
We first run a simulation to solve the relaxed-decoding problem in an environment with a very low probability error. We do this to start building a policy library for the PPR algorithm. 
We train the agent from scratch to solve the environment using a DDQN. From \figureautorefname{\ref{fig:Env0}} we see that the agent starts converging around $6000^{th}$ episode. The policy generated after the completion of $10000$ episodes will be added to the policy library to be used in harder environments.

\begin{figure}[t]
\centering
\begin{subfigure}[b]{1\linewidth}
   \includegraphics[trim=0 -0 0 0,angle=0,width=1\linewidth]{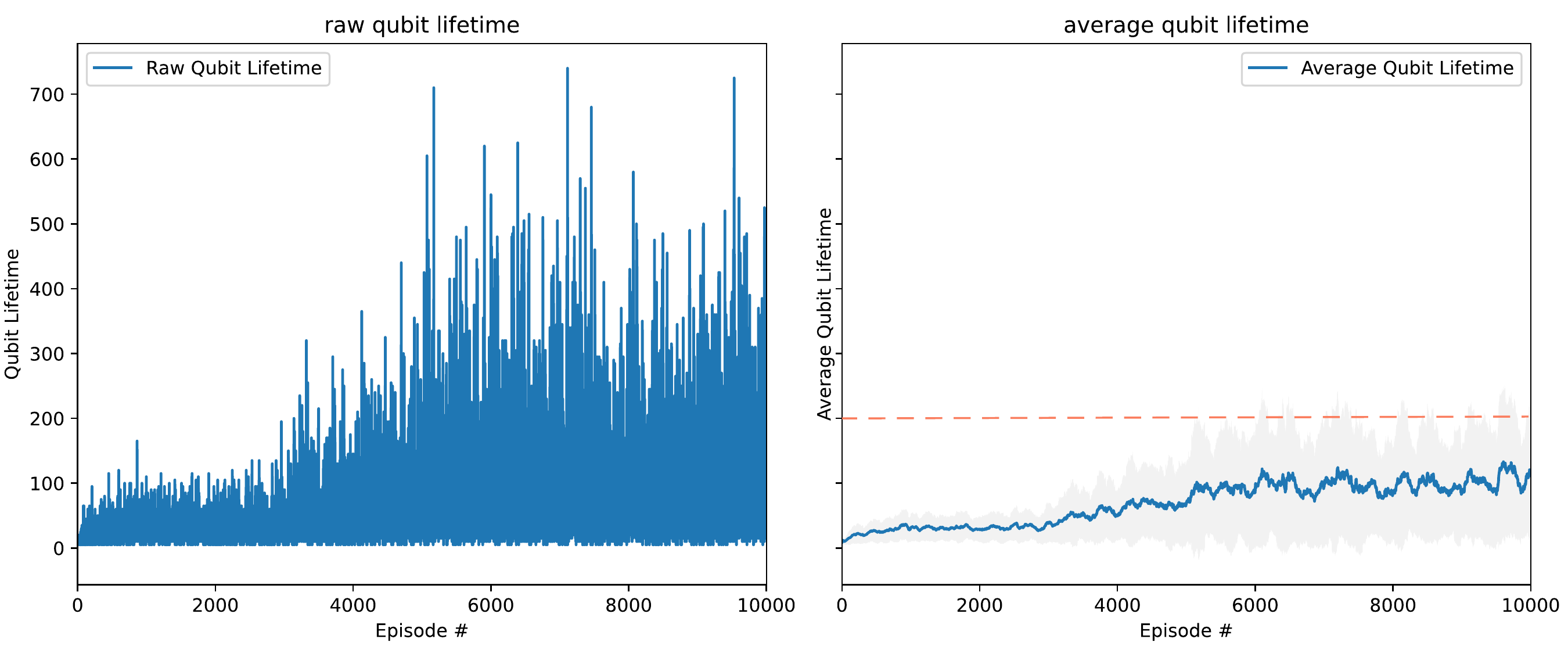}
   \subcaption{Training from scratch simulation for \emph{Environment-2.}}
   \label{fig0.007scratch}
\end{subfigure}
\begin{subfigure}[b]{1\linewidth}
   \includegraphics[trim=0 -0 0 0,angle=0,width=1\linewidth]{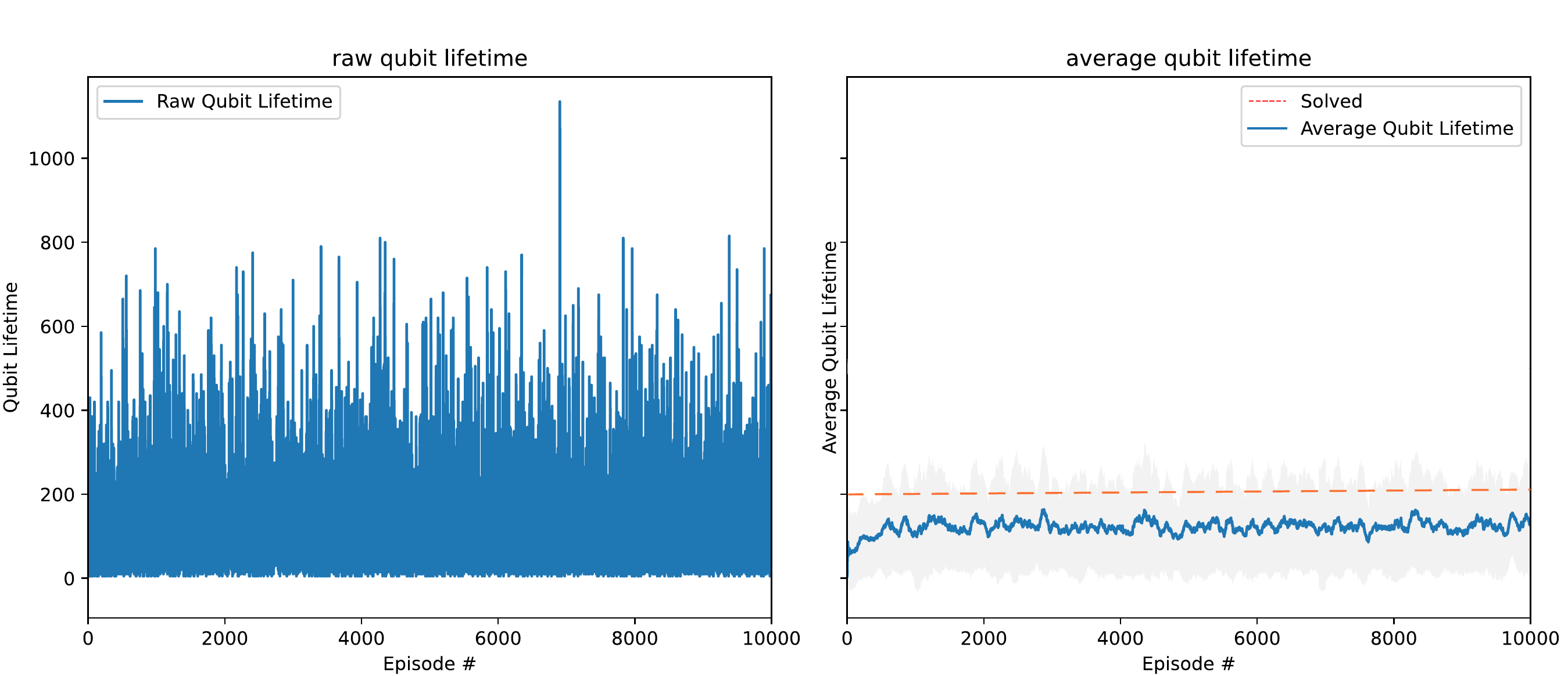}
   \subcaption{Policy reuse simulation for \emph{Environment-2}. Policy library: Policy trained from scratch for \emph{Environment-0} and policy trained for \emph{Environment-1} described in \ref{sec:0.005} .} 
   \label{fig0.007reuse}
\end{subfigure}
 \caption{(Color online) Training from scratch and policy reuse results for \emph{Environment-2}: Single bit-flip error rate and a measurement error rate of 0.007. }
\label{fig0.007}
\end{figure}

\subsection{\label{sec:0.005}\emph{Environment-1}: error-probability=0.005}
\begin{figure}
\centering
\captionsetup[subfigure]{justification=centering}
\begin{subfigure}[b]{0.45\linewidth}
\centering
\includegraphics[width=\linewidth]{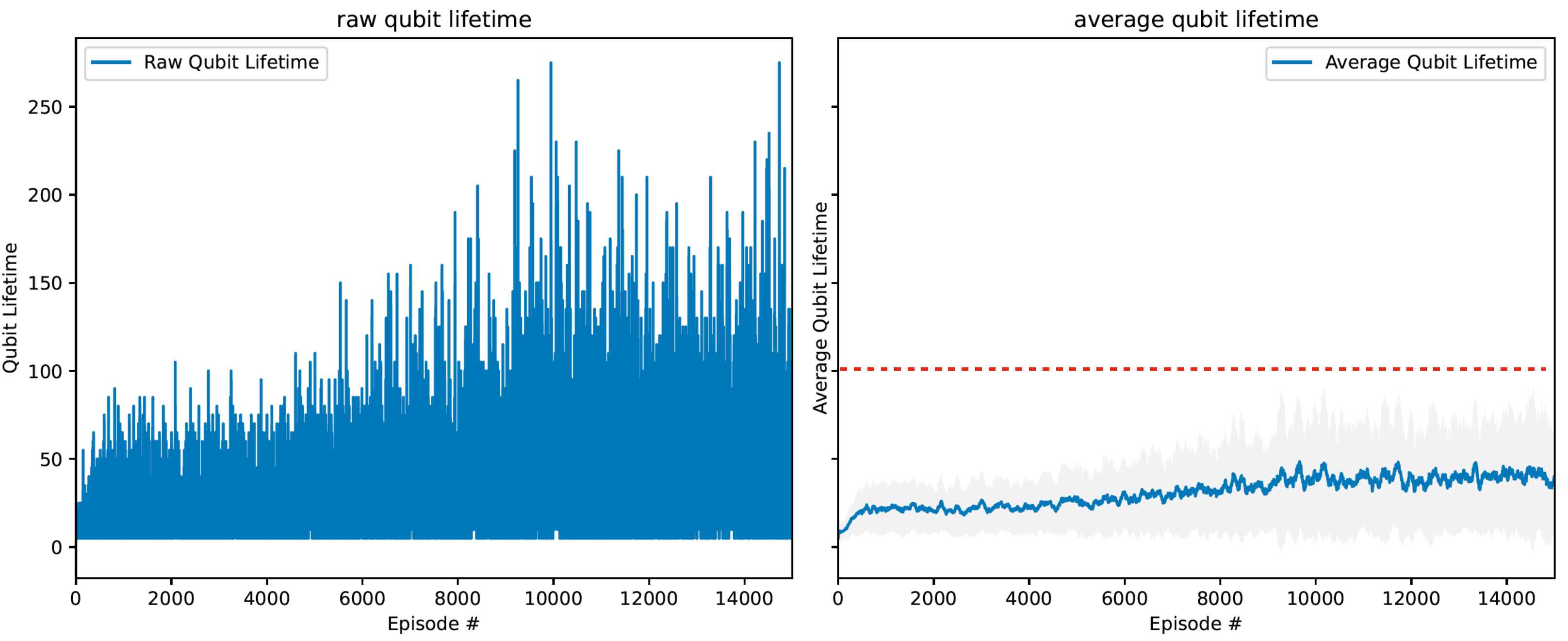}
   \subcaption{Training from scratch simulation for \emph{Environment-3}}
   \label{fig0.011scratch}
   \vspace{4ex}
\end{subfigure}
\quad
\begin{subfigure}[b]{0.45\linewidth}
\centering
   \includegraphics[width=\linewidth]{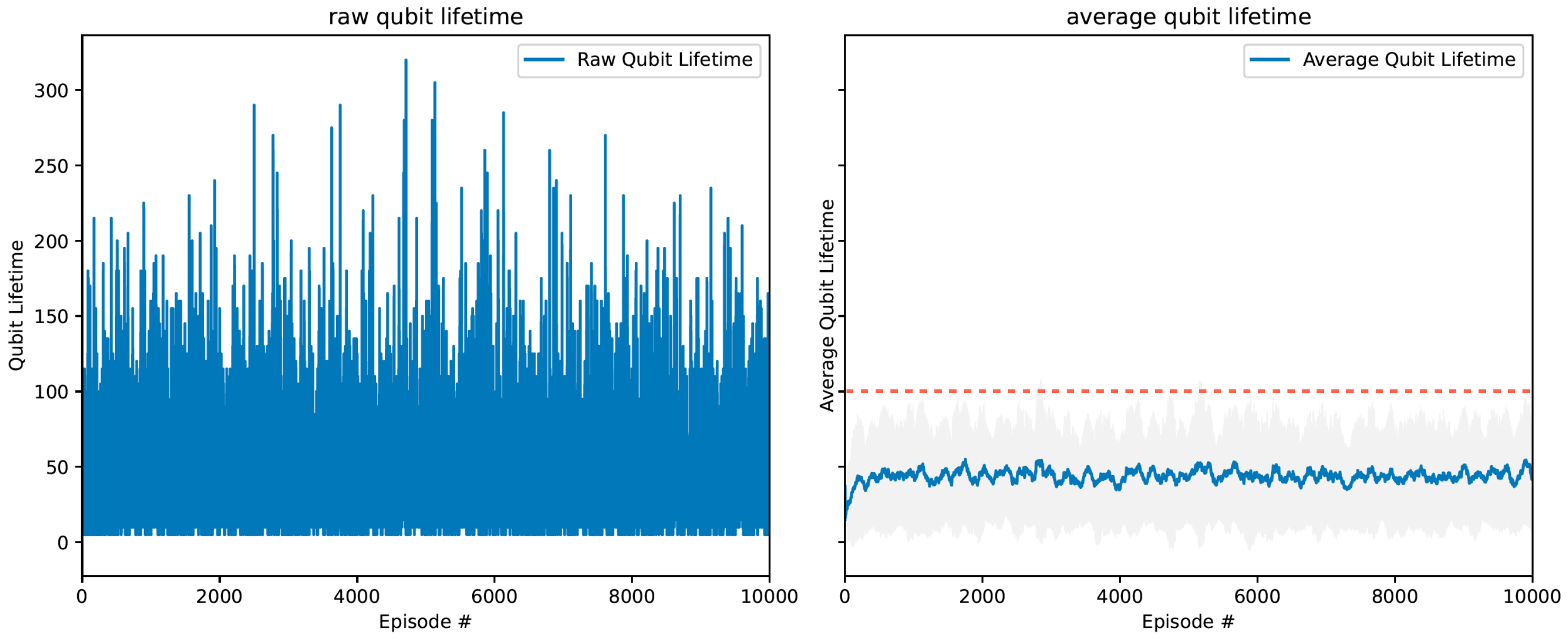}
   \subcaption{Policy reuse simulation for \emph{Environment-3}, using \emph{Environment-0} policy from the policy library.
   }
   \label{fig0.011_env0}
   \end{subfigure}
\quad
\begin{subfigure}[b]{0.45\linewidth}
\centering
   \includegraphics[width=\linewidth]{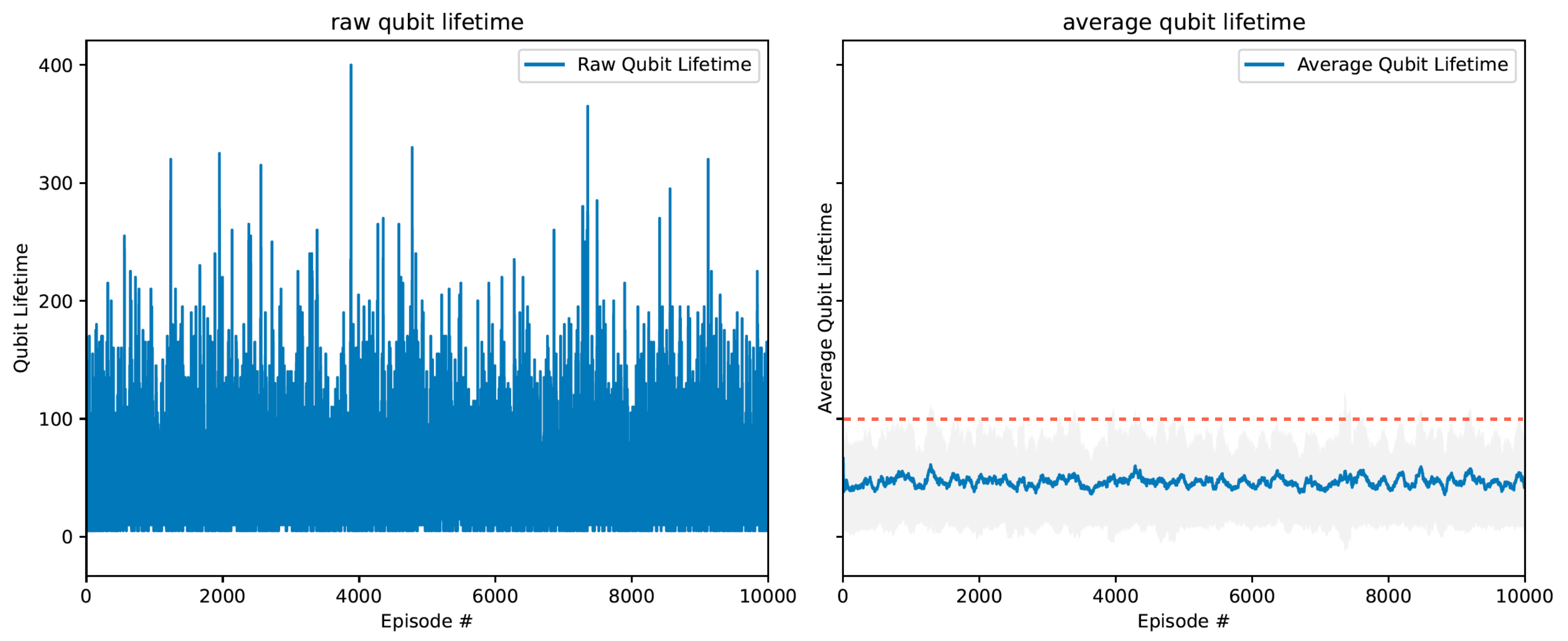}
   \subcaption{Policy reuse simulation for \emph{Environment-3}, using \emph{Environment-0}, \emph{Environment-1} policies from the policy library. }
   \label{fig0.011_env0_env1}
   \vspace{4ex}
\end{subfigure}
\quad
\begin{subfigure}[b]{0.45\linewidth}
\centering
   \includegraphics[width=\linewidth]{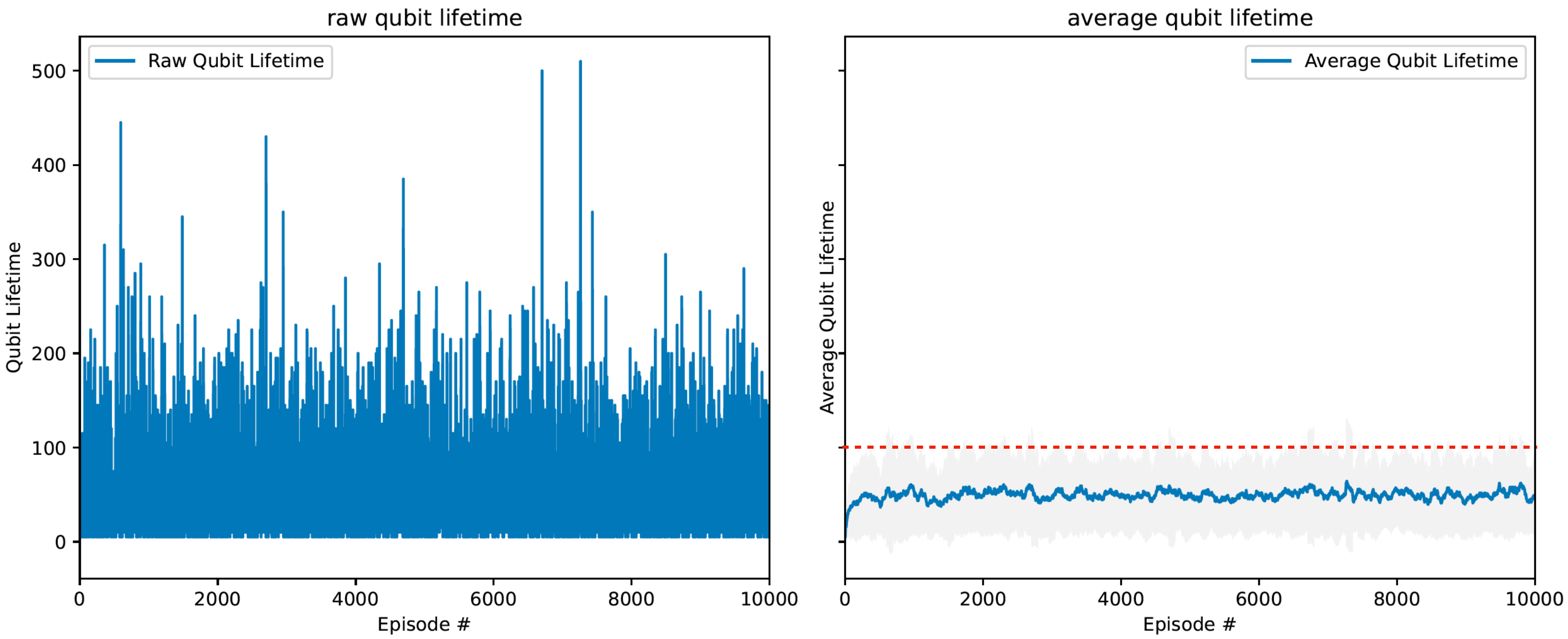}
   \subcaption{Policy reuse simulation for \emph{Environment-3}, using \emph{Environment-0}, \emph{Environment-1}, \emph{Environment-2}  policies from the policy library. }
   \label{fig0.011_env0_env1_env2}
   \vspace{4ex}
\end{subfigure}
 \caption{(Color online) Training from scratch and policy reuse results for \emph{Environment-3}: Single bit-flip error rate and a measurement error rate of 0.011. We observe that as the number of policies increases the agent is able to attain higher qubit lifetime, as well as converges faster.}
\label{fig0.011}
\end{figure}
Next, we apply the policies from \emph{Environment-0} to the next setting, \emph{Environment-1}, where the error-probability $p_{\rm{err}}$ is $0.005$. We compare the two simulations where we train the agent from scratch to solve \emph{Environment-1}, as well as using the PPR algorithm. In \figureautorefname{\ref{fig0.005scratch}} and \ref{fig0.005_env1}, we use the qubit lifetime of $500$ as the baseline as \cite{Sweke2020} is able to achieve an average qubit lifetime of $500$ using their model. 
In the training from scratch, the agent stabilizes around the $5000^{th}$ episode (\figureautorefname{\ref{fig0.005scratch}}), whereas the PPR algorithm as shown in \figureautorefname{\ref{fig0.005_env1}} achieves a higher qubit lifetime and stabilizes around the $3000^{th}$ episode. Additionally, when we compare the raw qubit lifetime in \figureautorefname{\ref{fig0.005scratch}} and \ref{fig0.005_env1}, we see that the PPR algorithm achieves a score greater than 2000, which is not achieved when we do training from scratch. 
\begin{figure}[t]
\centering
    \begin{subfigure}[t]{1\textwidth}
    \centering
    \includegraphics[width=1\textwidth]{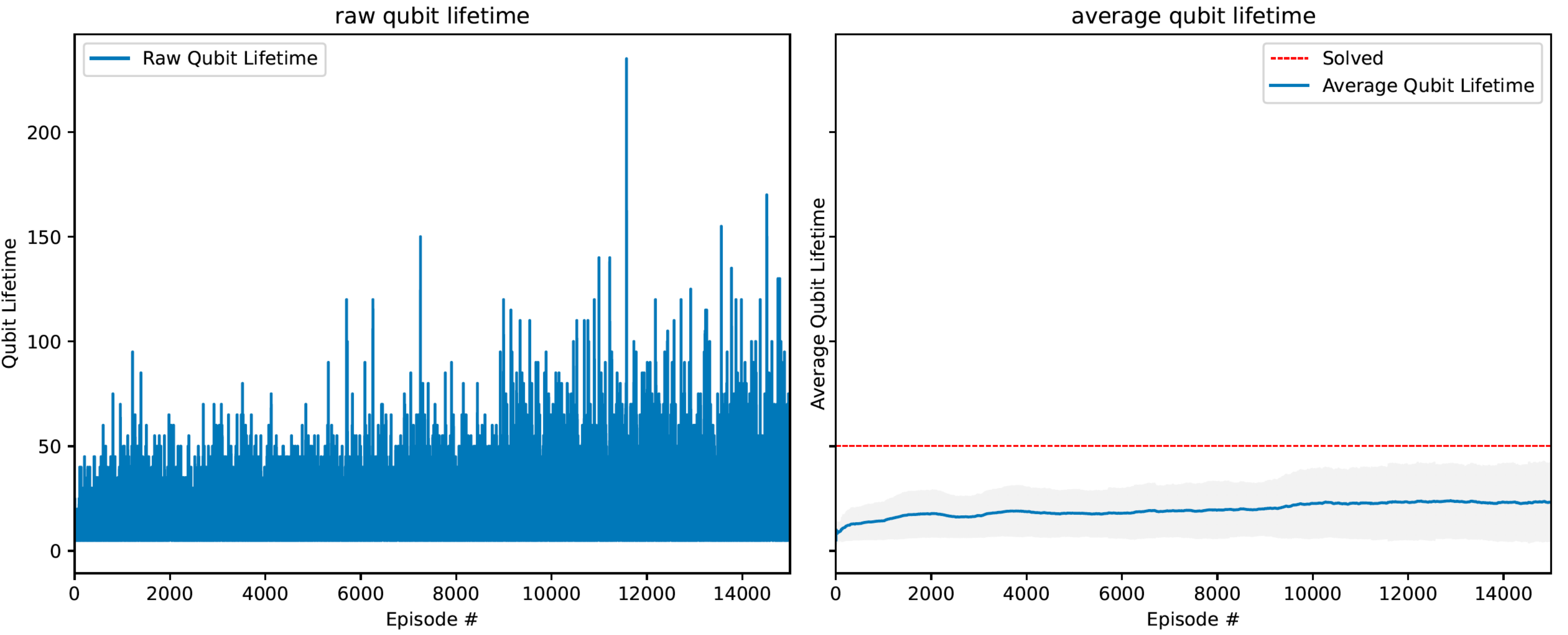}
       \subcaption{Training from scratch simulation for \emph{Environment-4}}
       \label{fig0.015scratch}
    \end{subfigure}
    \quad
    \begin{subfigure}[t]{1\textwidth}
    \centering
    \includegraphics[width=1\textwidth]{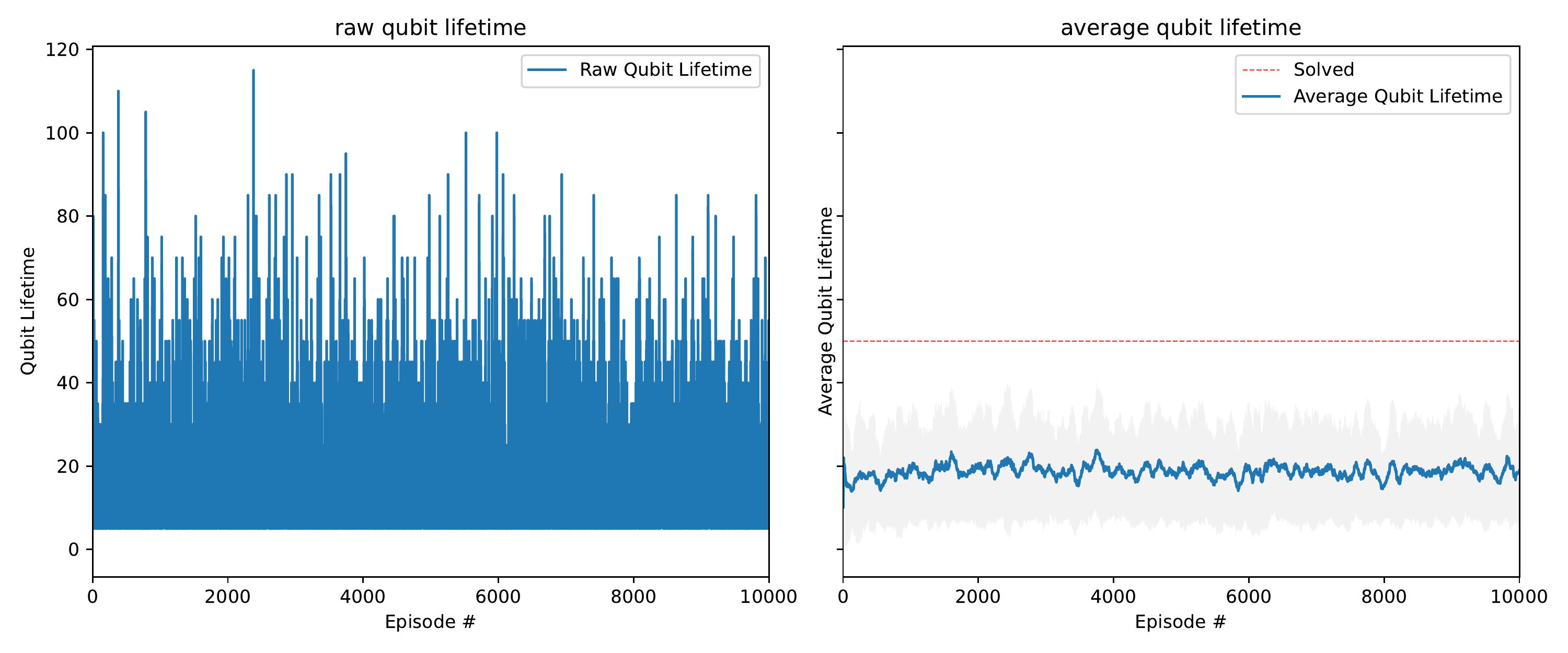}
       \subcaption{Policy reuse simulation for \emph{Environment-4} using \emph{Environment-3} policy from the policy library.}
    \end{subfigure}
\caption{(Color online) Training from scratch and policy reuse results for \emph{Environment-4}: Single bit-flip error rate and a measurement error rate of 0.015. }\label{fig5}
\end{figure}

\subsection{\label{sec:0.007}\emph{Environment-2}: error-probability=0.007}
In this environment, we increase the error-probability to $0.007$, which is relatively more challenging compared to the previous environments. We observe from \figureautorefname{\ref{fig0.007scratch}}, that the agent struggles to suppress the errors, and unlike the previous environment, the agent does not stabilize until $5000^{th}$ episode. However, when we implement the previous policy libraries (\emph{Environment-0}, \emph{Environment-1}), the agent starts learning around $2000^{th}$ episode and stabilizes well, as observed in \figureautorefname{\ref{fig0.007reuse}}.

\subsection{\label{sec:0.011}\emph{Environment-3}: error-probability=0.011}
The error-probability for this environment increases to $0.011$, making it more difficult for the agent to suppress errors. We see from \figureautorefname{\ref{fig0.011scratch}} that the training from scratch takes a longer time to converge/learn (14000 total episodes). From the results of the PPR algorithm, we see that the learning is relatively more stable even when a single policy is used, as shown in \figureautorefname{\ref{fig0.011_env0}}. In \figureautorefname{\ref{fig0.011_env0_env1}}, \ref{fig0.011_env0_env1_env2} there is a drastic improvement in the average-qubit lifetime as well as the convergence of the agent when we use two, and three policies. We note that the average-qubit lifetime is not very high, but it is understandable as the environment is difficult to solve due to the increase in the probability of error. 

\subsection{\label{sec:0.015}\emph{Environment-4}: error-probability=0.015}
\begin{figure}[t]
\captionsetup[subfigure]{justification=centering}
\centering
  \begin{subfigure}[]{0.5\linewidth}
    \centering
    \includegraphics[width=0.95\linewidth]{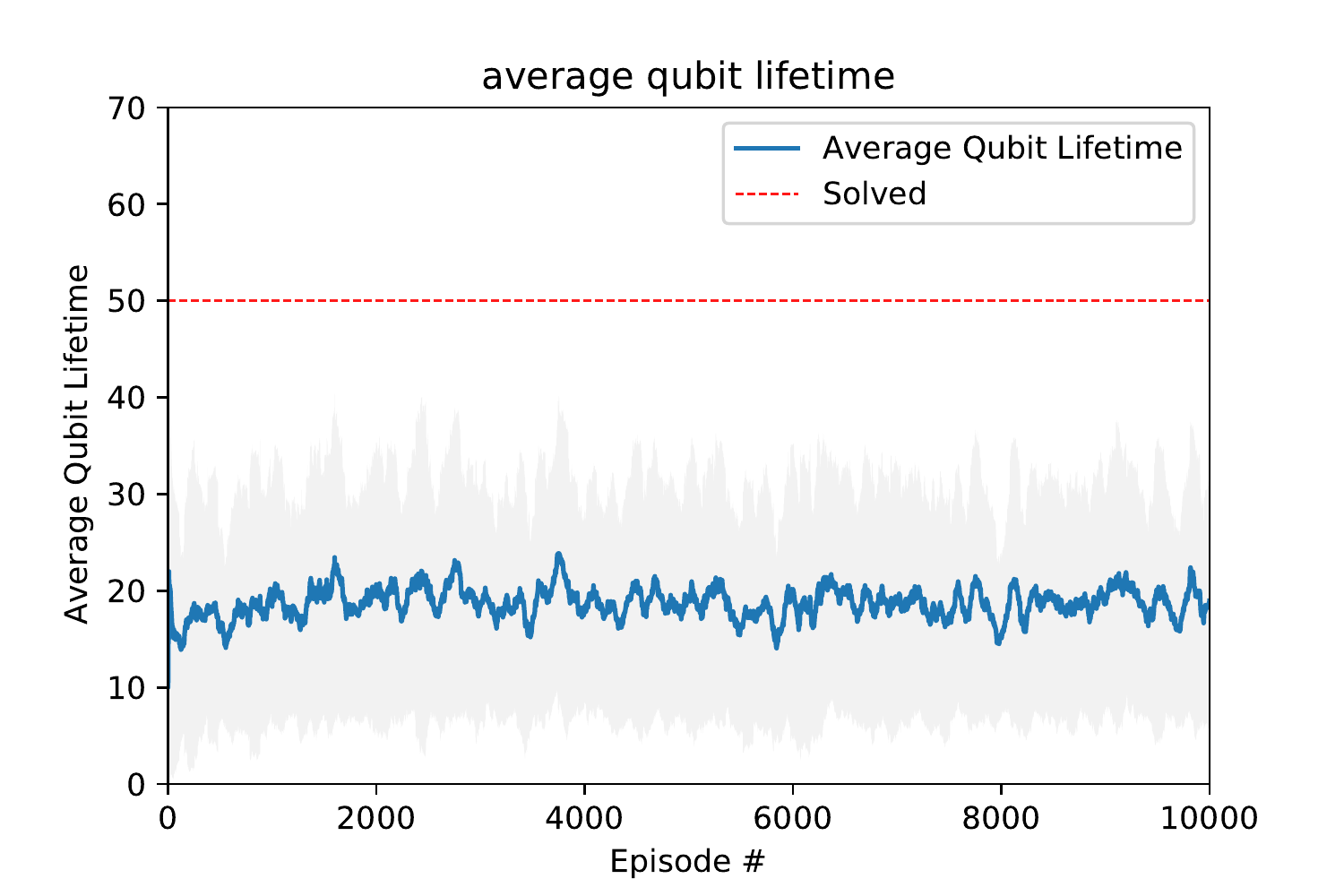} 
    \subcaption{One Policy: [0.011]} 
    \label{fig6a} 
    \vspace{2ex}
  \end{subfigure}
  \begin{subfigure}[]{0.5\linewidth}
    \centering
    \includegraphics[width=0.95\linewidth]{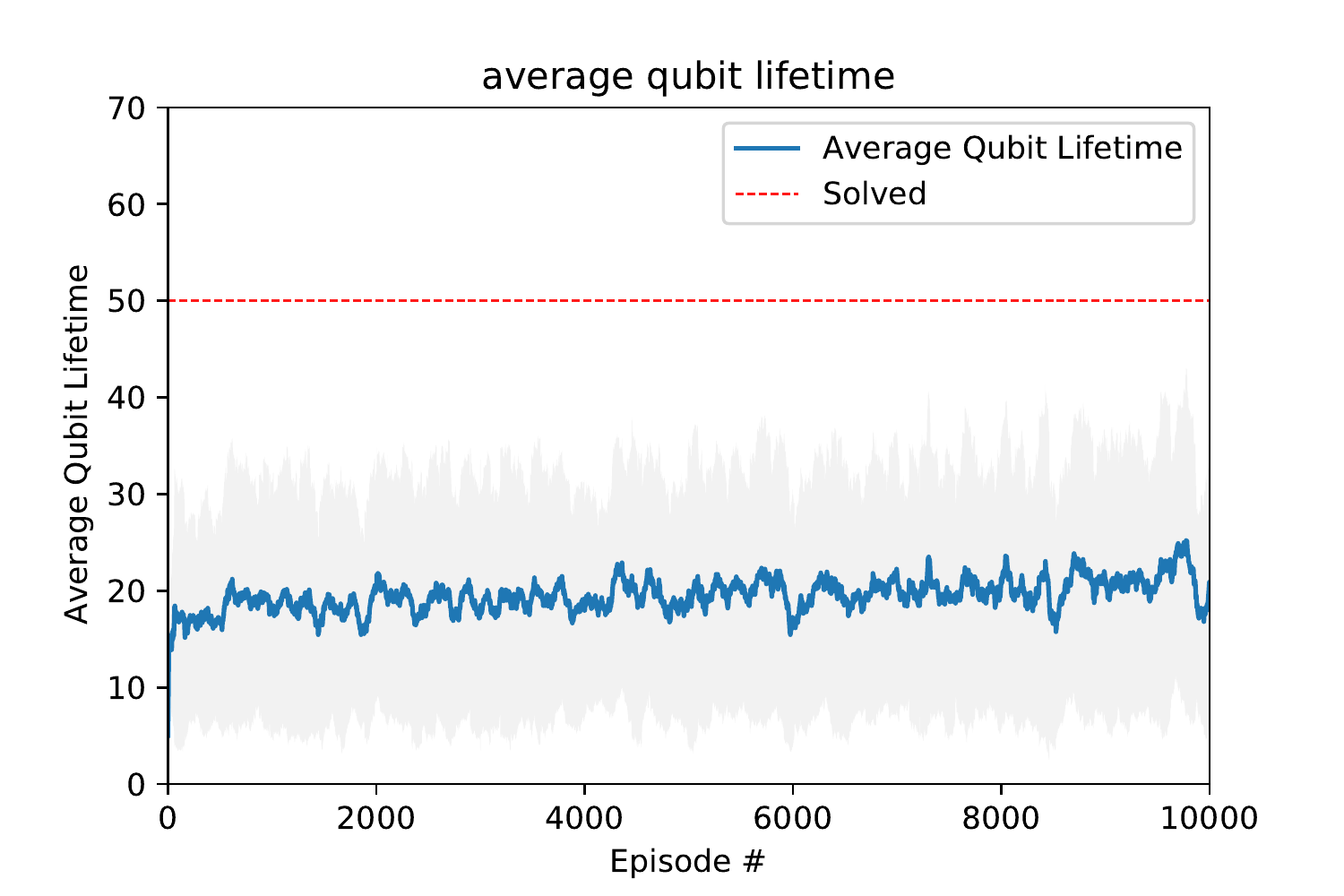}
    \subcaption{Two Policies: [0.007, 0.011]}
    \label{fig6b} 
    \vspace{2ex}
  \end{subfigure} 
  \begin{subfigure}[]{0.5\linewidth}
    \centering
    \includegraphics[width=0.95\linewidth]{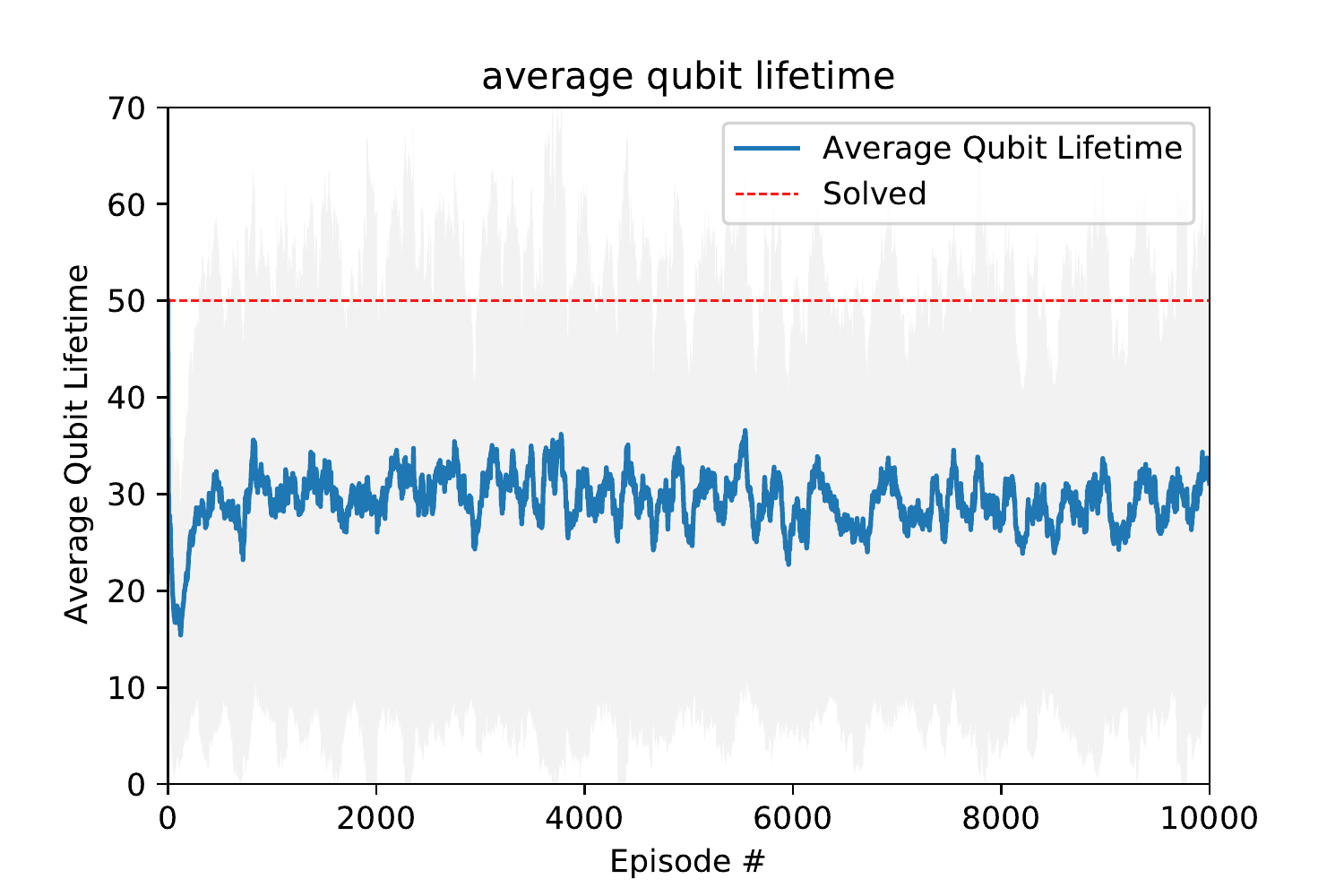} 
    \subcaption{Three Policies: [0.003, 0.007, 0.011]} 
    \label{fig6c} 
    \vspace{2ex}
  \end{subfigure}
  \begin{subfigure}[]{0.5\linewidth}
    \centering
    \includegraphics[width=0.95\linewidth]{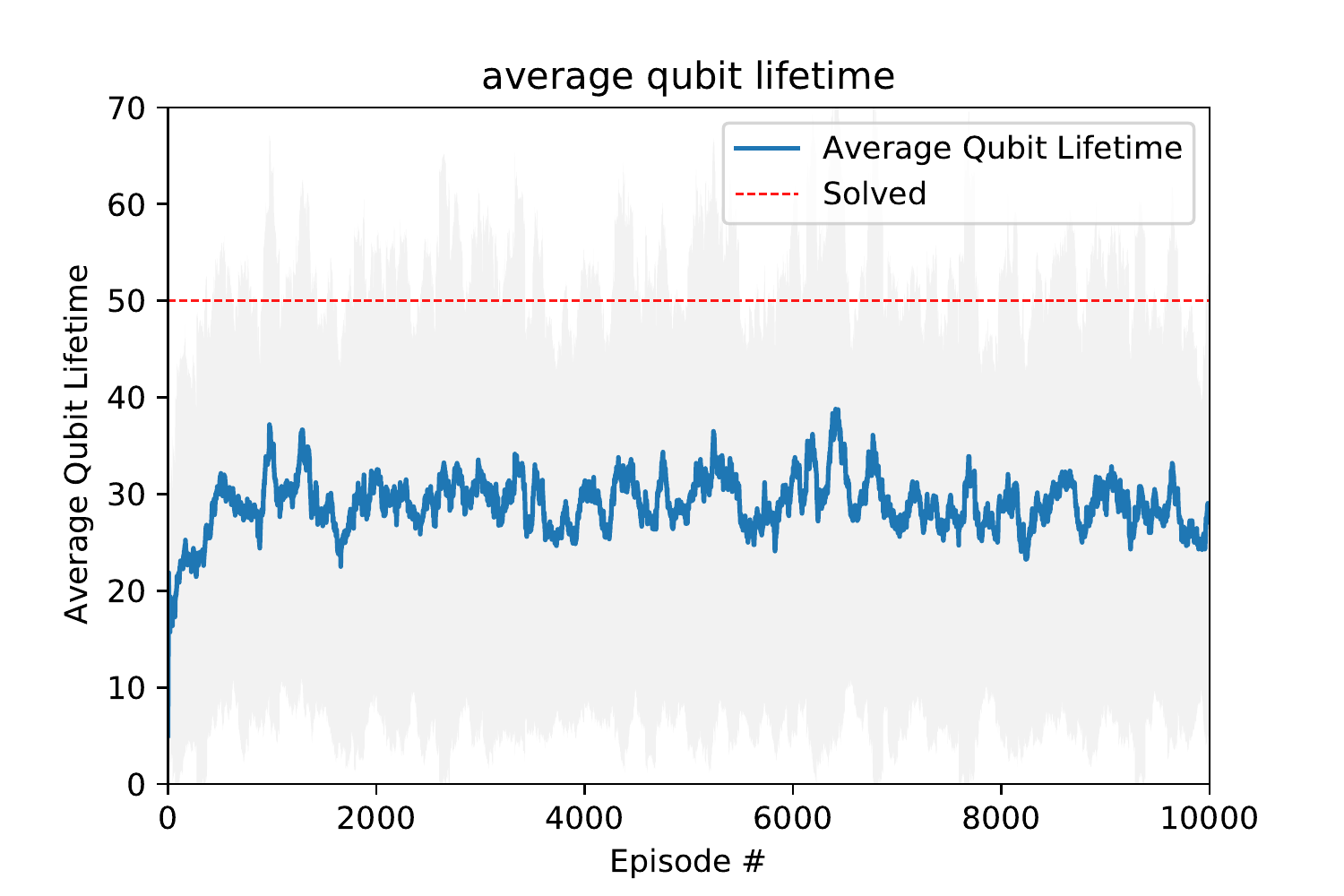}
    \subcaption{Four Policies: [0.003, 0.005, 0.007, 0.011]}  \label{fig6d}
    \vspace{2ex}
  \end{subfigure}
  \caption{(Color online) Policy reuse simulation for $p_{\textrm{err}}=0.015$ with different number of policies. We observe that there is a huge improvement in the \emph{average} qubit lifetime for three policies against two policies. There is not much notable difference when we increase the number of policies from three to four. However, we will show in \figureautorefname{\ref{fig:hist}} that the raw qubit lifetime increases for a higher number of policies.}
  \label{fig6} 
\end{figure}

The error-probability for this environment increases to $0.015$, it is the hardest environment we consider in this work. We see from \figureautorefname{\ref{fig5}} that single policy reuse of \emph{Environment-3} does not provide any improvement over training from scratch. In \figureautorefname{\ref{fig6}} we compare the results for using a different number of policies from the policy library. We observe that there is an immense improvement in the average qubit lifetime from two policies to three policies. However, the performance of the agent does not change much from three policies to four policies.  


\subsection{Different Number of Policies}
\begin{figure}[h]
\captionsetup[subfigure]{justification=centering}
\centering
  \begin{subfigure}[]{0.5\linewidth}
    \centering
    \includegraphics[width=0.95\linewidth]{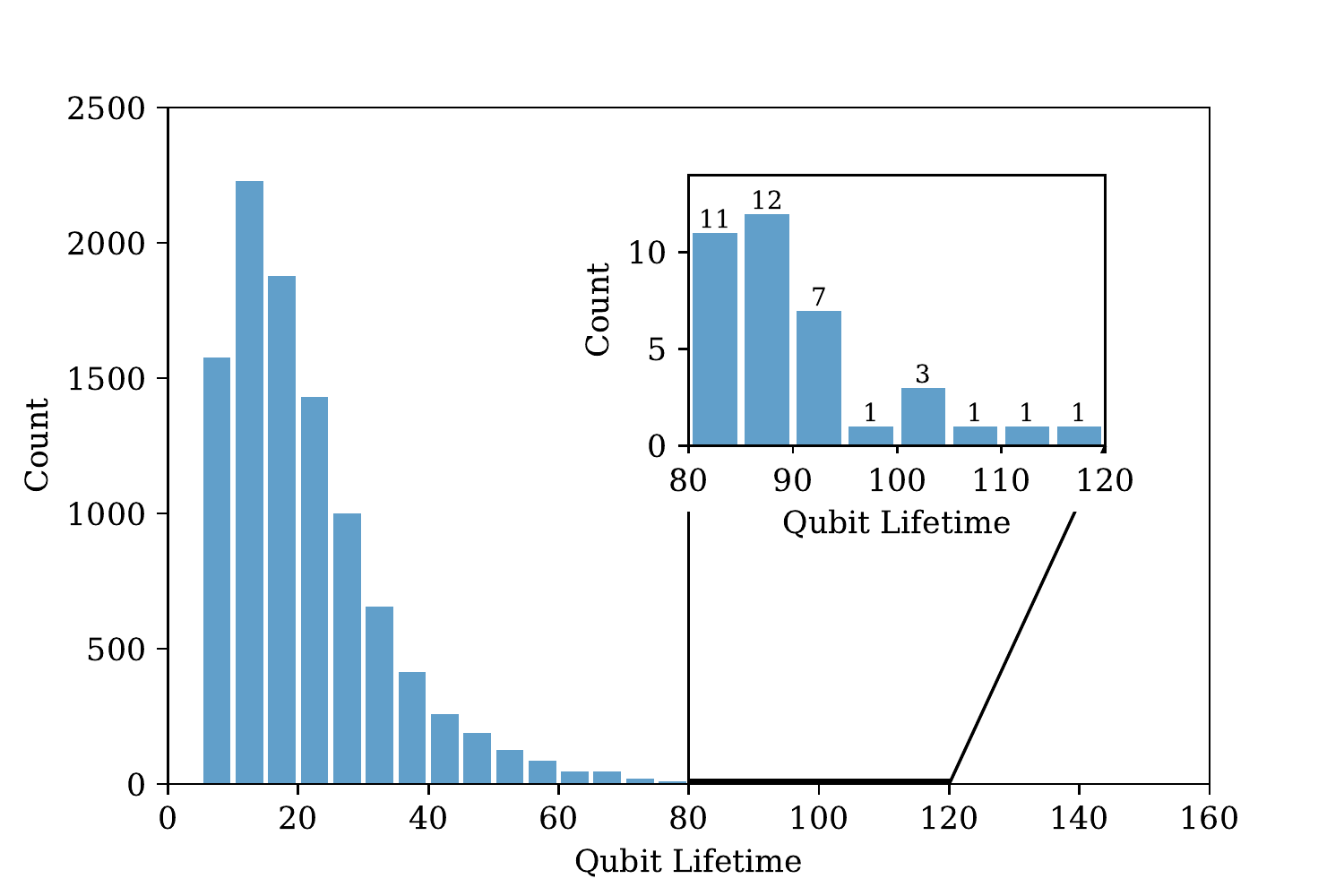} 
    \subcaption{\footnotesize Policy reuse simulation for \emph{Environment-4} using \textbf{one policy}: the \emph{Environment-3} policy. Within the inset, the qubit lifetimes falling between \emph{80} and \emph{120} are few and heavily skewed to the right.} 
    \label{fig:hista} 
    \vspace{2ex}
  \end{subfigure}
  \begin{subfigure}[]{0.5\linewidth}
    \centering
    \includegraphics[width=0.95\linewidth]{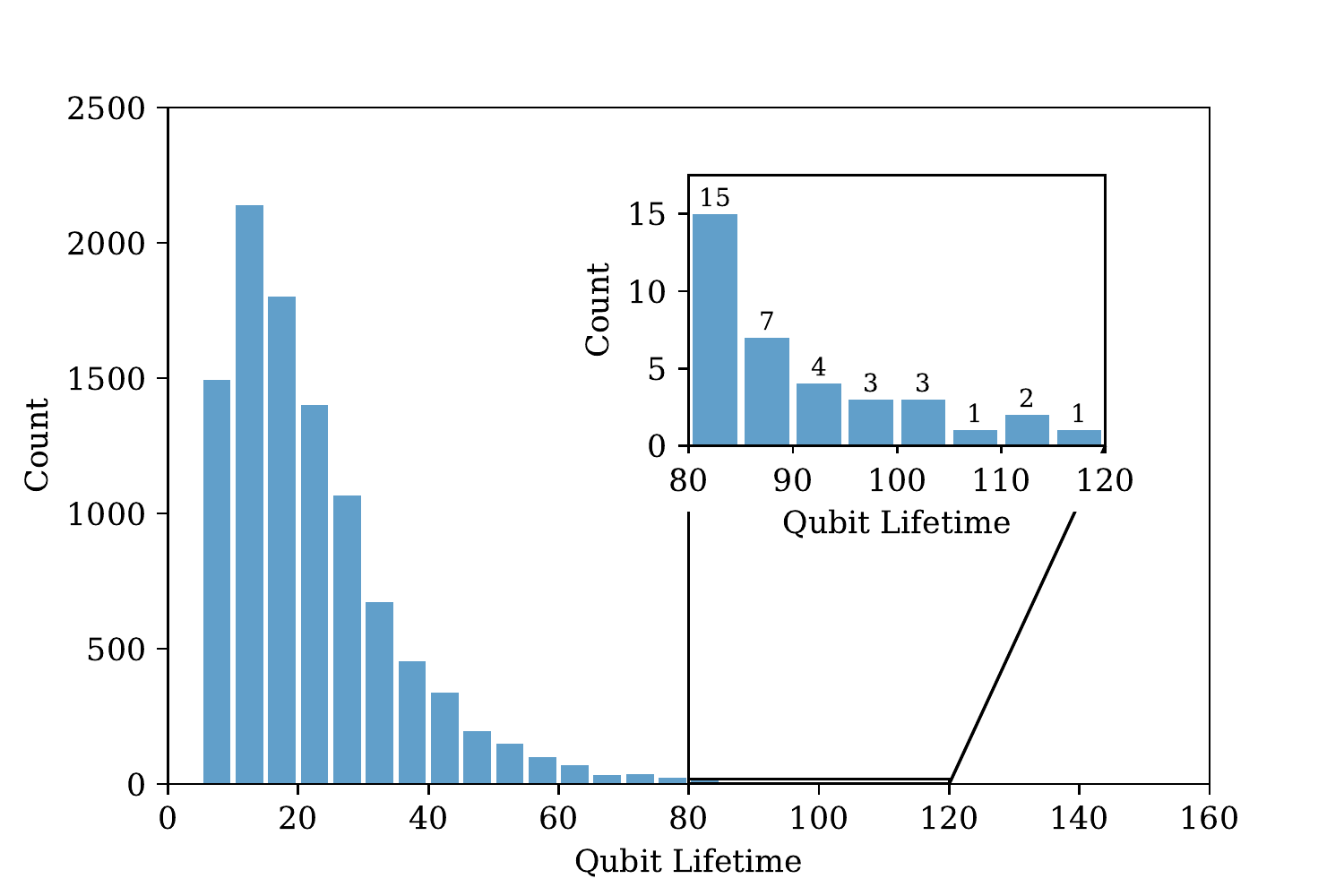} 
    \subcaption{\footnotesize Policy reuse simulation for \emph{Environment-4} using \textbf{two policies}: the \emph{Environment-2} \& \emph{Environment-3} policies. Within the inset, there is more occupation in the higher lifetimes compared to (a).} 
    \label{fig:histb} 
    \vspace{2ex}
  \end{subfigure} 
  \begin{subfigure}[]{0.5\linewidth}
    \centering
    \includegraphics[width=0.95\linewidth]{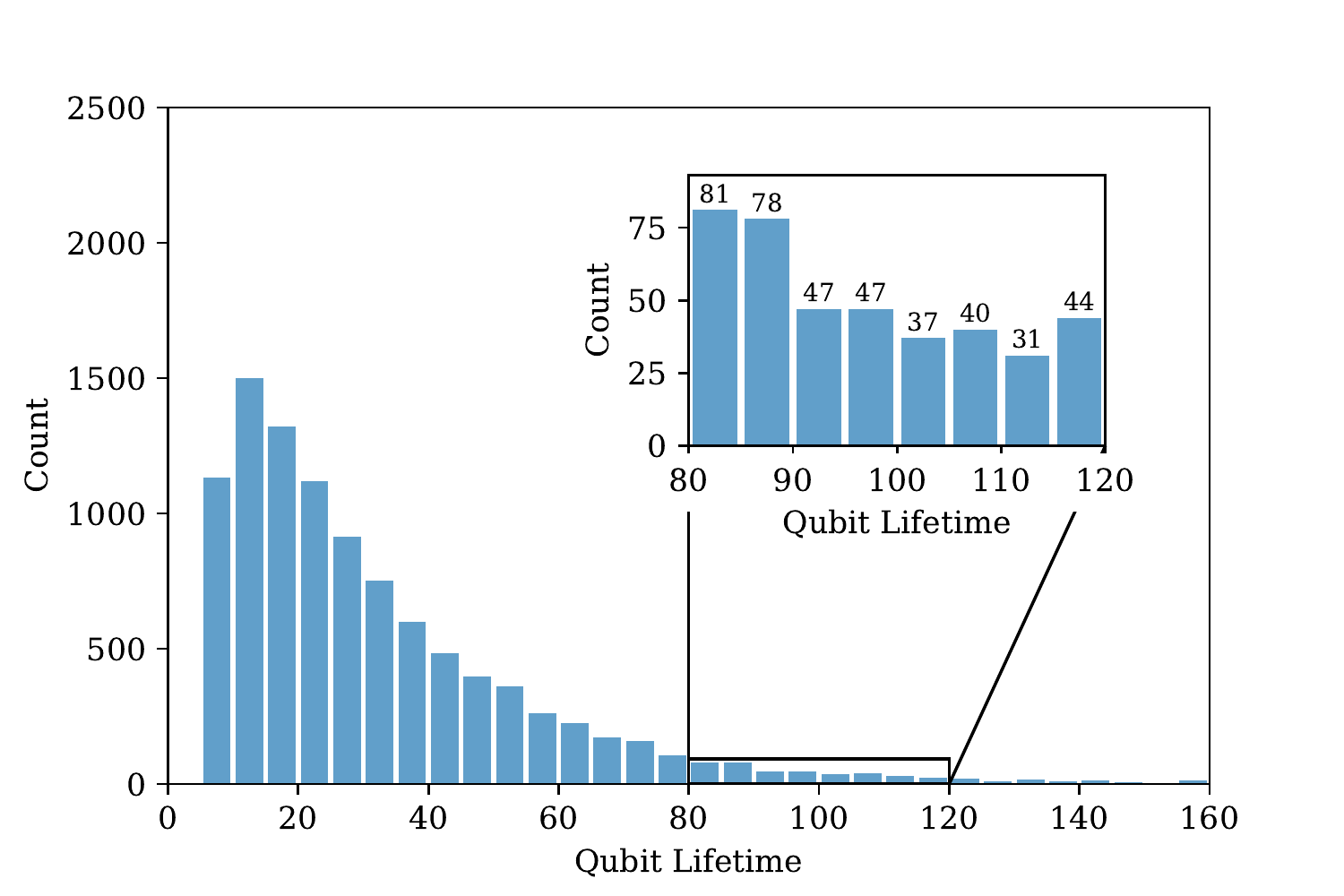} 
    \subcaption{\footnotesize Policy reuse simulation for \emph{Environment-4} using \textbf{three policies}: the \emph{Environment-0}, \emph{Environment-2}, \& \emph{Environment-3} policies. Within the inset, there is a large overall increase in occupation for lifetimes between \emph{80} and \emph{120} compared to (a) and (b).}
    \label{fig:histc} 
    \vspace{2ex}
  \end{subfigure}
  \begin{subfigure}[]{0.5\linewidth}
    \centering
    \includegraphics[width=0.95\linewidth]{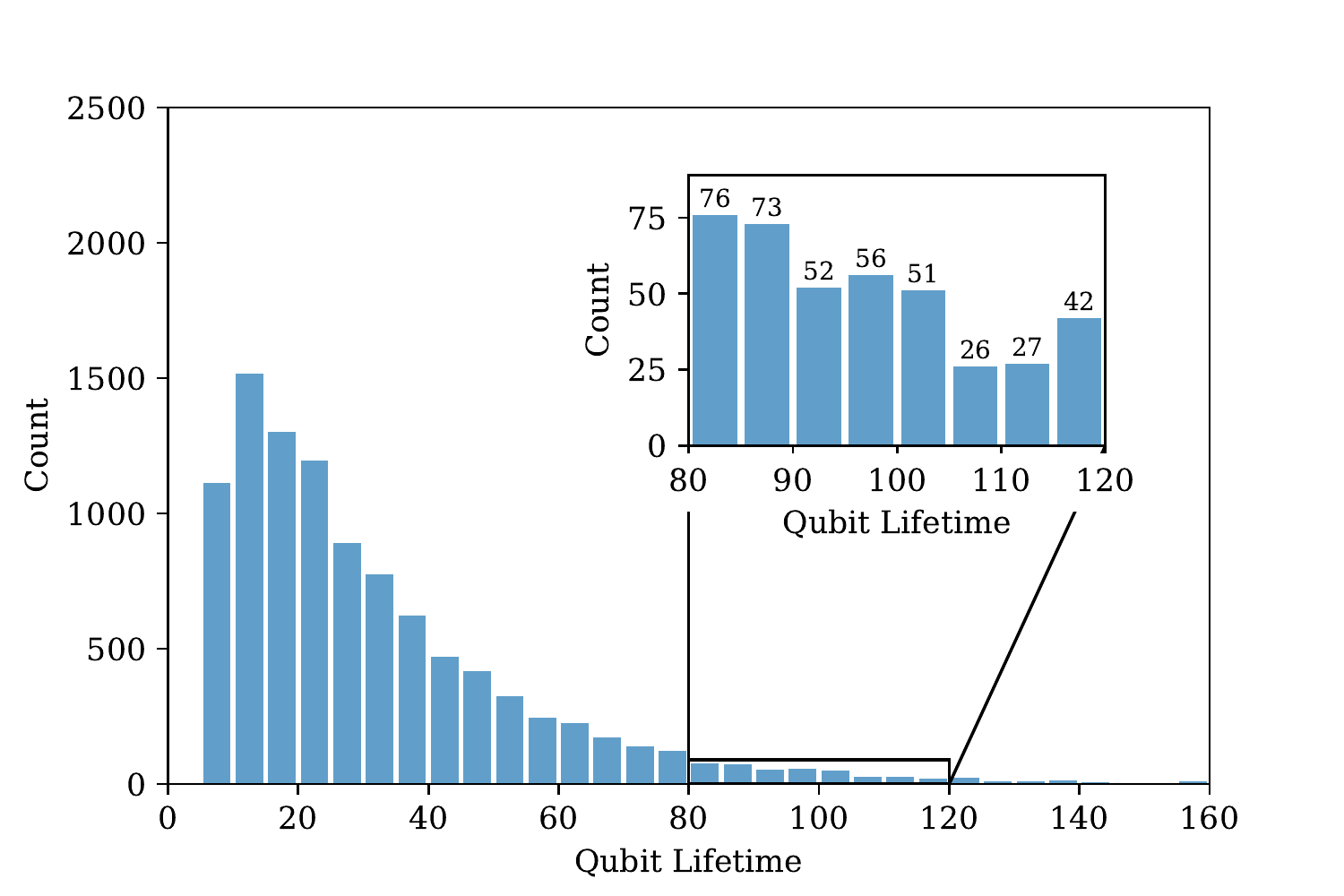} 
    \subcaption{\footnotesize Policy reuse simulation for \emph{Environment-4} using \textbf{four policies}: the \emph{Environment-0}, \emph{Environment-1}, \emph{Environment-2}, \& \emph{Environment-3} policies. Within the inset, there is more occupation in the higher lifetimes compared to (c).}  \label{fig:histd}
    \vspace{2ex}
  \end{subfigure}
  \caption{(Color online) Histograms for the policy reuse simulations of \emph{Environment-4} using different policies from the policy library. Each histogram plots the frequency of qubit lifetimes over the 10,000 episodes of each simulation. We observe that the simulations with more policies noticeably increased the qubit lifetimes of each episode as shown with higher occupation in the larger lifetime bins.}
  \label{fig:hist} 
\end{figure}
In \figureautorefname{\ref{fig:hist}} we illustrate the results of the algorithm in \emph{Environment-4} ($p_{err} = 0.015$) in the form of histograms. Each histogram represents the culmination of $10000$ episodes over a single simulation and plots the frequency of qubit lifetimes gathered at the termination of each episode. The histogram insets highlight the frequencies of lifetimes ranging between 80 and 120. As we can see from \figureautorefname{\ref{fig:hist}}, there is a dramatic increase from using 1-2 policies (\figureautorefname{\ref{fig:hista}}, \ref{fig:histb}) to 3-4 policies (\figureautorefname{\ref{fig:histc}}, \ref{fig:histd}). This increase demonstrates the algorithm's capabilities in lengthening the qubit lifetime for a larger number of episodes as more policies are loaded into the policy library.

In \tableautorefname{\ref{table:2}} we compare the average qubit lifetime of the logical qubit decoded by the agent trained from scratch and using policy reuse against the average qubit lifetime of a single faulty qubit. For the bit-flip noise explored in this work, we find that for $p_{\rm{err}}<0.011$ the decoding agent of the surface code $d=5$ logical qubit is able to attain an average qubit lifetime that is greater than that of the single faulty qubit. We attribute this to various factors. First, our PPR algorithm was not executed over multiple values of hyper-parameters. Therefore, the hyper-parameters used are not necessarily the optimal values. Second, we hope to see better performance when using more complex NN architectures \cite{Julien2022,esslinger2022deep,Wang2018}. Finally, as pointed out before, the increasing number of policies can increase the qubit lifetime. Therefore, building a more exhaustive policy library for different values of $p_{\rm{err}}$ will most likely improve the qubit lifetimes. 

\setlength{\tabcolsep}{0.5em} 
\begin{table}[!ht]
\centering
\begin{tabular}{|c|c|c|c|}
\hline
p\_{\rm{err}} &
  \begin{tabular}[c]{@{}c@{}}Single Faulty \\ Qubit Lifetime\end{tabular} &
  \begin{tabular}[c]{@{}c@{}}DDQN (training from \\ scratch) Surface Code\end{tabular} &
  \begin{tabular}[c]{@{}c@{}}Policy Reuse DDQN \\ Surface Code\end{tabular} \\ \hline
0.003 & 333 & 1000 &    ---              \\ \hline
0.005 & 200       & 250  & 400 (1 policy)   \\ \hline
0.007 & 143 & 120  & 180 (2 policies) \\ \hline
0.011 & 91  & 45   & 50 (3 policies)  \\ \hline
0.015 & 67  & 25   & 40 (4 policies)  \\ \hline
\end{tabular}

\caption{Table comparing the lifetime for a single faulty qubit against the average qubit lifetime obtained from training the agent from scratch (using DDQN) and policy reuse. The qubit lifetime for training from scratch surpasses that of the faulty qubit up to $p_{\rm{err}}=0.005$. The policy reuse is able to attain a higher average qubit lifetime compared to the faulty qubit up to $p_{\rm{err}}=0.007$. For higher $p_{\rm{err}}$ the policy reuse is not able to surpass the faulty qubit lifetime. However, the policy reuse results can be improved if we can tune the hyper-parameters, and/or increase the number of episodes for training the agent.} 
\label{table:2}
\end{table}
\section{\label{sec:Discussion}Discussion}

\subsection{Relevant Works}
\label{sec:RelevantWorks}
Machine learning (ML) techniques have been applied to tackle certain quantum computing challenges recently. Notable examples are quantum architecture search (QAS) \cite{fosel2021quantum,kuo2021quantum,ye2021quantum,ostaszewski2021reinforcement,chen2021quantum,yao2022monte,ding2022evolutionary,kimura2022quantum,duong2022quantum,he2022search,wang2022automated,chen2022generating,sogabe2022model}, quantum control \cite{sivak2022model,niu2019universal,bukov2018reinforcement,brown2021reinforcement,brown2022optimal}, quantum compiling \cite{moro2021quantum,zhang2020topological,he2021variational,pozzi2020using,chen2022efficient,chen2021quantum}, quantum error mitigation \cite{kim2020quantum} and quantum error correction \cite{nautrup2019optimizing,convy2022machine,andreasson2019quantum}.
The research fields mentioned above are all sequential decision-making tasks, common in quantum computing research at different levels of the quantum stack. All of those areas can benefit from the use of ML techniques, especially RL.

QAS is in the high-level part of the stack. The purpose of QAS is to find a quantum circuit architecture suitable for a particular task. The task might be generating a desired quantum state \cite{kuo2021quantum,ye2021quantum,kimura2022quantum}, finding an efficient circuit for solving chemical ground states \cite{ostaszewski2021reinforcement}, solving an optimization task \cite{yao2022monte,duong2022quantum}, optimizing a given quantum circuit for a particular hardware architecture \cite{fosel2021quantum} or performing a machine learning task \cite{ding2022evolutionary,duong2022quantum}. While our work's purpose is not to generate a quantum circuit, it does share some features with those QAS works. For example, the works \cite{kuo2021quantum,ye2021quantum} require the RL agent to decide suitable quantum operations to be added. It is similar to choosing a proper action to `correct' the errors after knowing the syndrome information. However, this environment can also be implemented  
Having the quantum circuit architecture, the next step is to compile the circuit into hardware-specific gate sequences. The compiled instructions may need to be further optimized or adjusted to be fault-tolerant or error-corrected. Our work is at this level. 
At the most fundamental level, the quantum control procedure is used to translate the gate sequences into control signals applied to quantum physical systems \cite{Krotov,GRAPE1,CRAB,GRAPE2,Riaz2019,GOAT}. These control signals (or pulses) are responsible for taking the qubit from one state to another. Optimal-control-based methods are widely used to improve the fidelity of quantum gate operations, by optimizing the control pulses\cite{brown2022optimal,Koch_2016}. In recent years, several open-source software packages have been developed to improve gate fidelity using optimal pulse control \cite{c3-optimize,qopt2021}. In \cite{matekole2022methods}, quantum optimal control of pulses has been executed on the experimental hardware and has shown an improvement in the performance of quantum logic gates over the default logic gates.
Quantum optimal control is an ideal problem for RL-based methods because it steers the quantum system to a particular target state by synthesizing the control fields for a given set of constraints, noise, and time. Deep-RL techniques have been implemented on the experimental quantum hardware of IBM and have shown some robustness to the errors\cite{sivak2022model,niu2019universal,bukov2018reinforcement,brown2021reinforcement,He2021,qctrl2021}.

%
The proposed continual RL framework is applied to surface code decoding, however, it is not limited to QEC decoding only. 
For example, in quantum compiling, we need to translate the high-level quantum algorithm or circuit architecture into a gate sequence suitable for particular quantum hardware and take into account the available quantum operation/gate sets or hardware topology 
If we can leverage the knowledge of compiling policies trained on other quantum computers with similar but different topologies or configurations, it will reduce the overall development timeline. Another field where our methods can be applied to is quantum control. Since the device noises usually drift and we may not fully capture the dynamics accurately, it is very difficult to derive an exact analytical model for the system Hamiltonian which is required to derive the control signals. 
If we can treat the device at each time step as a black box and train RL agents with the PPR algorithm as described in this paper, it may help in building the \emph{autonomous control signal calibrator}.
%
\subsection{Running on a real Quantum Computer}
%
The ultimate goal of investigating these AI/ML methods for quantum error correction or other relevant topics mentioned in \sectionautorefname{ \ref{sec:RelevantWorks}} is to actually run on quantum computers. The major challenges, which motivated this research, are the changing device noise patterns in many aspects. In addition, it is impractical to deploy a large number of training episodes on real quantum computers to capture the error pattern. However, researchers and engineers collect valuable information about the machine of interest and people can design approximate noise models to simulate such systems to some extent. Such information will also benefit the actual deployment of our proposed methods. The idea is that we can train RL agents against these approximate noise models and gain some knowledge of the system. If we can observe the system for a bit longer time, we may also collect information on how these noise drift and construct a series of training environments as the \emph{curriculum} to train the agent. After training with the curriculum, the agent has a better starting point to be trained with a real quantum computer with different noise patterns. For example, the agent may collect a library of policies to use when encountering real-world noise and this gained knowledge would significantly reduce the required training episodes.

%
\subsection{Hardware Accelerators and Model Distillation}
The practical application of the proposed method is to be working with real quantum hardware. One of the crucial issues is the latency of the inference process (e.g. generating suitable correcting actions in real time). This challenge can be potentially solved via specialized acceleration hardware such as a field-programmable gate array (FPGA) which is an integrated circuit designed to be configured or reprogrammed after manufacturing. 
Indeed, recent works have shown that FPGA can be used to accelerate the ML inference processes in certain fields such as computer vision \cite{westby2021fpga} and particle physics data analysis \cite{duarte2019fpga}. 
It is expected that such techniques can be applied in RL as well \cite{cho2019fa3c}.
%
Consider a step further, we can even compress the trained models for more efficient inference. For example, existing methods such as model distillation \cite{polino2018model,gou2021knowledge} can be used to largely reduce the model size while keeping the performance to the desired level. 
In addition, trained models in the policy library can be distilled into several crucial policies. For example, environments may be different but still share many common features. The policies trained for these similar environments should be similar. It is desirable to extract the most important information from these policies to form \emph{eigenpolicies} \cite{fernandez2006probabilistic}. Various methods have been proposed to distillate the knowledge learned by the RL agent to build a more efficient learning scheme \cite{rusu2015policy,traore2019discorl}. We expect those advanced continual RL techniques can be applied in the context of quantum computing.
\subsection{Beyond Decoding}
RL and other ML techniques have been applied to QEC decoding problems \cite{Torlai2017,krastanov2017deep,varsamopoulos2017decoding}. 
However, the performance of a QEC scheme is largely determined by the encoding method. The proposed method can be used to study encoding problems as well, e.g. finding optimal code structures under various noise patterns. For example, it is possible to utilize the framework in this paper to study how to apply a continual RL agent to find optimal quantum error correction codes when the device error changes over time. For example, in \cite{nautrup2019optimizing} the RL agent was exposed to a more complicated surface code geometry, and the goal was to find the optimal connectivity of the surface code instead of just having nearest-neighbor connectivity on a simple square lattice. The agent was able to find unique qubit connectivities that yielded the best surface codes. 

Another interesting application of RL that goes beyond decoding is \emph{autonomous quantum error correction}, where the agent's task is to find the optimal encoding which is robust to the noisy dynamics of the quantum system. The authors in \cite{wang2022automatedQEC} introduced an algorithm called AutoQEC that achieves this task while maximizing the fidelity of the logical qubit and were also able to discover a new quantum error correction code. Moreover, in \cite{Florian2018} an ab initio method was proposed where machine learning was employed for a full QEC protocol discovery. Even though this method works for a small system of qubits, it provides an important outlook on the flexibility of RL that is capable of discovering QEC along with error mitigation, from scratch.

%
\section{\label{sec:Conclusion}Conclusion}
In this work, we demonstrate the capabilities of the DDQN-PPR model in decoding the surface code when bit-flips have occurred. Through simulations in quantum environments with varying noise levels, we numerically show there is an increase in the agent's performance when it utilizes PPR to apply the knowledge from previously learned policies to a new noise environment. The agent's improvement is reflected by the increase in the average qubit lifetime seen with the PPR simulations when compared to the simulations done from scratch. While this algorithm has shown its capabilities in addressing QEC, the framework detailed here is general enough to be applied to other aspects of quantum computing such as QAS and QOC, where noise also plays a prominent role.

\clearpage
\begin{acknowledgments}
This work is supported by the U.S.\ Department of Energy, Office of Science, Office of High Energy Physics program under Award Number DE-SC-0012704, Office of Workforce Development for Teachers and Scientists (WDTS) under the Science Undergraduate Laboratory Internships Program (SULI) \& BNL High School Research Program (HSRP) and the Brookhaven National Laboratory LDRD \#20-024. This research used resources of the Oak Ridge Leadership Computing Facility, which is a DOE Office of Science User Facility supported under Contract DE-AC05-00OR22725. This research used resources of the National Energy Research Scientific Computing Center (NERSC), a U.S. Department of Energy Office of Science User Facility located at Lawrence Berkeley National Laboratory, operated under Contract No. DE-AC02-05CH11231 using NERSC award HEP-ERCAPm4138.

\end{acknowledgments}
\par\vfill\break

\appendix
\section{Appendixes}
\label{sec:Appendix}
\advance\vsize by 2cm 
\advance\voffset by -1cm
\begin{algorithm}[H]
\begin{algorithmic}
\State \textbf{Define} the transition pair to be in the form $(s_{t}, a_{t}, r_{t}, s_{t+1})$
\State \textbf{Given} an environment, $\mathcal{E}$
\State \textbf{Given} the replay memory, $\mathcal{D}$
\State \textbf{Given} a new policy, $L_0$
\State \textbf{Given} the maximum number of episodes to execute, $K$
\State \textbf{Given} the maximum number of steps per episode, $H$
\State \textbf{Initialize} the rewards list $W$
\For{episode $=1,2,\ldots,K$} 
\State Reset the environment $\mathcal{E}$
    \For{$t = 1,2,\ldots,H$}
    \State Select action $a_{t}$ from $L_0$ using greedy policy selection
    \State Perform the action $a_{t}$
    \State Observe the new state $s_{t +1}$
    \State \textbf{if} not done:
        \State \quad Assign the next state in the transition pair to be the new state $s_{t+1}$
    \State \textbf{else}:
        \State \quad Assign the next state in the transition pair to be none
    \State Store the transition pair to $\mathcal{D}$ and assign current state $s$ to be next state $s_{t+1}$
    \State Optimize $L_0$
    \State \textbf{if} done:
    \State \quad Update $W$
    \State \quad break
\EndFor
\EndFor
\State \textbf{Return} mean($W$), new policy $L_0$
\end{algorithmic}
\caption{$Q$-learning algorithm}
\label{q_learn_alg}
\end{algorithm}
\par\vfill\break
\advance\vsize by -2cm 
\advance\voffset by 1cm 

\par\vfill\break 
\advance\vsize by 2cm 
\advance\voffset by -1cm 
\scalebox{0.9}{
\begin{minipage}{\linewidth}
\begin{algorithm}[H]
\begin{algorithmic}
\State \textbf{Define} the transition pair to be in the form $(s_{t}, a_{t}, r_{t}, s_{t+1})$
\State \textbf{Given} an environment, $\mathcal{E}$
\State \textbf{Given} the replay memory, $\mathcal{D}$
\State \textbf{Given} a past policy to try $L_i$ and the new policy $L_0$
\State \textbf{Given} the maximum number of episodes to execute, $K$
\State \textbf{Given} the maximum number of steps per episode, $H$
\State \textbf{Given} the parameters $\psi$ and $\nu$
\State \textbf{Initialize} the rewards list $W$
\For{episode $=1,2,\ldots,K$} 
\State Reset the environment $\mathcal{E}$
    \For{$t$ $=1,2,\ldots,H$}
    \State Sample a random probability $p$
    \State \textbf{if} $p <= \psi$
    \State \quad Select action $a_{t}$ from $L_i$ using greedy policy selection
    \State \textbf{else}:
    \State \quad Select action $a_{t}$ from $L_0$ using greedy policy selection
    \State Perform the action $a_{t}$
    \State Observe the new state $s_{t+1}$
    \State Update $\psi$ to be $\psi \cdot \nu$
    \State \textbf{if} not done:
        \State \quad Assign the next state in the transition pair to be the new state $s_{t+1}$
    \State \textbf{else}:
        \State \quad Assign the next state in the transition pair to be none
    \State Store the transition pair to $\mathcal{D}$ and assign current state to be next state
    \State Optimize $L_0$
    \State \textbf{if} done:
    \State \quad Update $W$
    \State \quad break
\EndFor
\EndFor
\State \textbf{Return} mean($W$), new policy $L_0$
\end{algorithmic}
\caption{$\pi$-exploration algorithm}
\label{policy_reuse_pi_alg}
\end{algorithm}
\end{minipage}
}
\par\vfill\break
\advance\vsize by -2cm 
\advance\voffset by 1cm 

\par\vfill\break 
\advance\vsize by 2cm 
\advance\voffset by -1cm 
\scalebox{0.9}{
\begin{minipage}{\linewidth}
\begin{algorithm}[H]
\begin{algorithmic}
\State \textbf{Given} a new task $\Omega$ we want to solve
\State \textbf{Given} a policy library $L = \{\Pi_{1}, \cdots, \Pi_{n} \}$
\State \textbf{Given} an initial temperature parameter $\tau$ and the incremental size $\delta\tau$ for the Boltzmann policy selection strategy
\State \textbf{Given} the maximum number of episodes to execute, $K$
\State \textbf{Given} the maximum number of steps per episode, $H$
\State \textbf{Given} the parameters $\psi$ and $\nu$ for the $\pi$-exploration strategy
\State \textbf{Initialize} replay memory $\mathcal{D}$ to capacity $N$
\State \textbf{Initialize} a new policy $L_{0}$
\State \textbf{Initialize} a target policy $\hat{L}_{0} = L_{0}$
\State \textbf{Initialize} target network update rate, $T$  
\State \textbf{Initialize} the rewards $W_{i}$ to $0$
\State \textbf{Initialize} the number of episodes where policy $\Pi_{i}$ has been chosen, $U_{i} = 0, \forall i = 1, \cdots, n$
\For{episode $=1,2,\ldots,K$} 
\State Get probability vector $p$ from softmax function inputting $W$ and $\tau$
\State Sample from $p$ to get an index for the action policy $k$
    \State \textbf{if} {$k == 0$}:
        \State \quad Use the q-learning algorithm to get rewards $R$ and $L_0$
    \State \textbf{else}:
        \State \quad Use the $\pi$-exploration algorithm to get $R,\ L_0$
    \State Update $W_{k} \gets \frac{W_{k} \cdot U_{k} + R}{U_{k} + 1}$ 
    \State Update $U_{k} \gets U_{k} + 1$
    \State Update $\tau \gets \tau + \delta\tau$
    \State \textbf{if} episode is divisible by $T$:
        \State \quad Update $\hat{L}_{0}$ with $L_0$
\EndFor
\end{algorithmic}
\caption{Probabilistic Policy Reuse with double deep $Q$-learning}
\label{policy_reuse_alg}
\end{algorithm}
\end{minipage}
}
\advance\vsize by -2cm 
\advance\voffset by 1cm 

\par\vfill\break

\bibliographystyle{ieeetr}
\bibliography{fullbib}

\end{document}